\documentstyle[12pt]{article}

\catcode`@=11

\newtoks\@stequation

\def\subequations{\refstepcounter{equation}%
  \edef\@savedequation{\the\c@equation}%
  \@stequation=\expandafter{\theequation}
  \edef\@savedtheequation{\the\@stequation}
  \edef\oldtheequation{\theequation}%
  \setcounter{equation}{0}%
  \def\theequation{\oldtheequation\alph{equation}}}

\def\endsubequations{%
  \setcounter{equation}{\@savedequation}%
  \@stequation=\expandafter{\@savedtheequation}%
  \edef\theequation{\the\@stequation}%
  \global\@ignoretrue}

\def\citen#1{%
\if@filesw \immediate \write \@auxout {\string \citation {#1}}\fi
\@tempcntb\m@ne \let\@h@ld\relax \def\@citea{}%
\@for \@citeb:=#1\do {%
  \@ifundefined {b@\@citeb}%
    {\@h@ld\@citea\@tempcntb\m@ne{\bf ?}%
    \@warning {Citation `\@citeb ' on page \thepage \space undefined}}%
    {\@tempcnta\@tempcntb \advance\@tempcnta\@ne
    \setbox\z@\hbox\bgroup 
    \ifnum0<0\csname b@\@citeb \endcsname \relax
       \egroup \@tempcntb\number\csname b@\@citeb \endcsname \relax
       \else \egroup \@tempcntb\m@ne \fi
    \ifnum\@tempcnta=\@tempcntb 
       \ifx\@h@ld\relax 
          \edef \@h@ld{\@citea\csname b@\@citeb\endcsname}%
       \else 
          \edef\@h@ld{\hbox{--}\penalty\@highpenalty
            \csname b@\@citeb\endcsname}%
       \fi
    \else   
       \@h@ld\@citea\csname b@\@citeb \endcsname
       \let\@h@ld\relax
    \fi}%
 \def\@citea{,\penalty\@highpenalty\hskip.13em plus.1em minus.1em}%
}\@h@ld}
\def\@citex[#1]#2{\@cite{\citen{#2}}{#1}}%
\def\@cite#1#2{\leavevmode\unskip
  \ifnum\lastpenalty=\z@\penalty\@highpenalty\fi
  \ [{\multiply\@highpenalty 3 #1
      \if@tempswa,\penalty\@highpenalty\ #2\fi 
    }]\spacefactor\@m}
\catcode`@=12

\newcommand{\beq}{\begin{equation}}
\newcommand{\eeq}{\end{equation}}
\renewcommand{\theequation}{\thesection.\arabic{equation}}

\begin{document}
\begin{flushright}
ITP-SB-92-56
\end{flushright}

\vskip .5in

\begin{center}
{\Large \bf {Construction of Modular Branching Functions from
Bethe's Equations in the 3-State Potts Chain}}
\vskip 1cm

{\large{Rinat Kedem\footnote{rinat@max.physics.sunysb.edu}}}

{\large{Barry M. McCoy\footnote{mccoy@max.physics.sunysb.edu}}}

\vskip .5cm
Institute for Theoretical Physics\\
State University of New York \\
Stony Brook, NY 11794

\vskip .5cm
\end{center}
\begin{abstract}
We use the single particle excitation energies and the completeness
rules of the 3-state anti-ferromagnetic Potts chain, which have been
obtained from Bethe's equation, to compute the modular invariant
partition function. This provides a fermionic construction for the
branching functions of the $D_4$ representation of $Z_4$
parafermions which complements the previous bosonic constructions. It
is found that there are oscillations in some of the correlations and a
new connection with the field theory of the Lee-Yang edge is
presented.
\end{abstract}

\section{Introduction}
The theory of integrable quantum spin chains was initiated by Bethe in
1931~\cite{Bethe} in his study of the spin 1/2 Heisenberg
anti-ferromagnet. From that beginning, particularly in the last 20
years, an enormous number of one dimensional quantum spin systems have
been discovered which, along with their two dimensional statistical
counterparts, have the remarkable property that their energy
eigenvalues are given by the solutions of a system of equations which
have become known as Bethe's equations:
\beq
(-1)^{M+1}\left[{\sinh(\lambda_j-iS\gamma)\over{\sinh(\lambda_j+iS\gamma)}}
\right]^{N}=\prod_{k=1}^{L}{\sinh(\lambda_j-\lambda_k-i\gamma)\over
{\sinh(\lambda_j-\lambda_k+i\gamma) }}.
\eeq
Here $M$ is the number of sites in the chain, $N$ and $L$ are related
to $M$ (typically $N=M$ or $2M$), and $S$ and $\gamma$ are parameters
which characterize the specific models.

One of the features derived from these Bethe's equations is that in
the limit $M\rightarrow \infty$ the spectrum of lowlying excitations
$E_{ex}$ above the ground state is expressed in terms of a set of
single particle levels $e_{\alpha}(P)$ depending on a momentum $P$ and
combined with a set of rules as
\beq
E_{ex}-E_{GS}=\sum_{\alpha,rules} e_{\alpha}(P_{i}),
\eeq
\beq
P=\sum_{\alpha,rules}P_{i}^{\alpha},
\eeq
and almost without exception one of the rules of combination is a
``fermi'' exclusion rule:
\beq
P_{i}^{\alpha}\neq P_{j}^{\alpha}\quad {\rm if}\quad i\neq j.
\eeq
The form for energy levels (1.2) and (1.3) is referred to as a
quasi-particle spectrum.  Furthermore, in many of these spin chains one
or more $e_{\alpha}$ vanish as $P\rightarrow 0$
\beq
e(P)\sim v|P|,
\eeq
where $v$ is positive and is called the speed of sound.

Much more recently, in 1984, a powerful new formalism was
invented~\cite{BPZ} to study those integrable systems for which there
is no mass gap and (1.5) holds. This method, known as conformal field
theory, is more axiomatic. It deals with a continuum approximation to
the spin chain (or two dimensional statistical system) and instead of
starting from a Hamiltonian it starts from a symmetry principle such
as the Virasoro algebra, a Kac-Moody algebra, or modular invariance
which captures many of the essential features that lead to the
integrability of the systems described by Bethe's equation (1.1).

One of the main objects of computation in conformal field theory is
the partition function which is expressed by means of modular
invariance~\cite{Cardy,CIZ} in terms of Virasoro characters or (more
generally) branching functions $b_{j}(q)$ as
\beq
Z=\sum N_{k,l}b_k (q) b_l ({\bar q}).
\eeq
Here $q({\bar q})$ refers  to the right (left) moving excitations with
\beq
q,{\bar q}=e^{-{{2 \pi v}\over {M k_B T}}},
\eeq
where T is the temperature and $k_B$ is Boltzmann's constant.

The Virasoro characters and branching functions $b_{i}(q)$ are
solutions to the equations of modular
transformation~\cite{FF,RC,Kac1,KW}.  Their construction typically
starts with one or more free boson Fock spaces, and then excludes
certain null vectors. They are typically given by explicit formulas
with several powers of the product
\beq
 Q(q)=\prod_{n=1}^{\infty}(1-q^n)
\eeq
in the denominator, and a power series in q in the numerator, times a
fractional power which is usually written as $q^{-c/24 +h_k}$.  Here
the constant $c$ is referred to as the central charge, and $h_k$ are
known as the conformal dimensions.

The question now arises as to the relation between the solutions of
Bethe's equations and the results of conformal field theory. In
particular, one wants to compute the partition function (1.6) starting
from Bethe's equations (1.1) (or related functional equations).
Recently an important advance in this project was made by Kl{\"u}mper and
Pearce~\cite{KP1,KP2}, who computed the central charge and conformal
dimensions for the $A_{N+1}$ series of the $A_1^{(1)}$ models classified
by Pasquier~\cite{Pas}.  However the computation of the full character
expansion (1.6) and its relation to the quasi-particle energy spectrum
(1.2) is still lacking.

It our purpose here to complete this project and to compute the full
partition function (1.6) for a particular quantum spin model: The
anti-ferromagnetic 3-state Potts chain. In particular, we will show
that the partition function is constructed from the single particle
levels of (1.2) which satisfy the fermi exclusion rule (1.4). This
provides a physical interpretation of the model, which complements the
usual computation that starts with free bosons. The result of
conformal field theory, as obtained by Pearce~\cite{pearce2}, is that the
partition function of the spin system is
\beq
Z=\sum e^{-E_{n}/k_B T}=e^{-Me_0/k_B T}Z_{pf4},
\eeq
where $e_{0}$ is the ground state energy per site~\cite{Al1} and
$Z_{pf4}$ is the $D_4$ representation of the $Z_4$ parafermionic
partition function of~\cite{GQ}:
\beq
Z_{pf4}=[b_0^0(q)+b_4^0(q)][b_0^0(\bar q)+b_4^0(\bar q)]+
4b_2^0(q)b_2^0(\bar q)+ 2b_0^2(q)b_0^2(\bar q) +2b_2^2(q)b_2^2(\bar q),
\eeq
$b_m^{\ell}$  given in one of the equivalent bosonic forms of
(2.25), (2.27) and appendix B. We here obtain (1.10) starting from
Bethe's equation (1.1) for the finite lattice and obtain
new fermionic representations for $b_m^{\ell}$ given by
(3.13) and (3.19) for $b_0^0$, $b_2^0$, and$b_4^0$ in $Q=0$ and (4.16)
for $b_0^2$ and $b_2^2$ in $Q=\pm1$.

Our method is to combine the results of~\cite{Al1}, which derives the
spectrum of the anti-ferromagnetic 3-state Potts model of the form
(1.2), staring from the Bethe's equation derived by
Albertini~\cite{Al}, with the completeness study of~\cite{Al2} and the
finite size corrections of ~\cite{KP2}. In section~2 we summarize the
results of these papers which are needed here, as well as the
conformal field theory predictions for the model. In Section~3 we
compute the partition function in the channel $Q=0$ and in section~4
we do the same in the channel $Q=1$. We conclude in section~5 with a
discussion of the physical implication of our results in terms of what
we call an infrared anomaly. We also discuss the oscillations which
are predicted to occur in the correlation functions, and a connection
with the field theory~\cite{cardy2,isz} of the Lee-Yang edge~\cite{ly}
of the Ising model.

\section{Formulation}
\setcounter{equation}{0}
The 3-state anti-ferromagnetic Potts chain is specified by the
Hamiltonian
\beq
H={{2}\over{\sqrt3}}\sum_{j=1}^{M}\{ X_{j}+X_{j}^{\dagger} +
Z_{j}Z_{j+1}^{\dagger}+Z_{j}^{\dagger}Z_{j+1}\},
\eeq
where
\beq
 X_{j}=I\otimes I\otimes \cdots \otimes {\underbrace X_{j^{th}}}
 \otimes \cdots \otimes I,
\eeq
\beq
Z_{j}=I\otimes I \otimes \cdots \otimes {\underbrace Z_{j^{th}}}
 \otimes\cdots \otimes I.
\eeq
Here $ I$ is the $3 \times 3$ identity matrix, the elements of the $3
\times 3$ matrices X and Z are
\beq
X_{j,k}=\delta_{j,k+1}\quad({\rm mod}3),
\eeq
\beq
Z_{j,k}=\delta_{j,k}\omega^{j-1},
\eeq
\beq
\omega=e^{2\pi i/3},
\eeq
and we impose periodic boundary conditions $Z_{M+1}\equiv Z_{1}$.

This spin chain is invariant under translations and under spin
rotations. Thus the eigenvalues may by classified in terms of $P$, the
total momentum of the state, and $Q$, where $e^{2 \pi i Q/3}$ is the
eigenvalue of the spin rotation operator. Here $P=2 \pi n/M$ where $n$
is an integer $0\leq n \leq{M-1}$, and $Q=0,\pm 1$. Furthermore because
$H$ is invariant under complex conjugation there is a conserved $C$
parity of $\pm1$ in the sector $Q=0$ and the sectors $Q=\pm1$ are
degenerate.

This spin chain is integrable because of its connection with the
integrable 3-state Potts model of statistical mechanics. The
eigenvalues satisfy functional equations \cite{Al} \cite{Pearce} which
are solved in terms of a Bethe's equation (1.1)\cite{Al} with
\beq
N=2M, \quad\gamma=\pi/3,\quad S=1/4,
\eeq
and
\beq
L=2(M-|Q|)\quad {\rm for}\quad Q=0,\pm1.
\eeq
In terms of these ${\lambda_k}$, the eigenvalues of the transfer
matrix of the statistical model are
\beq
\Lambda(\lambda)=\left[{\sinh({\pi i\over6})\sinh({\pi i\over3})
\over{\sinh({\pi i\over4}-\lambda)\sinh({\pi i\over4}+\lambda)}}
\right]^{M}\prod_{k=1}^{L}{\sinh(\lambda-\lambda_k)\over{\sinh({\pi
i\over12}+\lambda_k)}},
\eeq
the eigenvalues of the Hamiltonian (2.1) are
\beq
E=\sum_{k=1}^{L}\cot(i\lambda_k+{\pi\over{12}})-{2M\over{\sqrt3}},
\eeq
and the corrsponding momentum is
\beq
e^{iP}=\Lambda(-i\pi/12)=\prod_{k=1}^{L}{\sinh(\lambda_k+
{\pi i\over12})\over{\sinh(\lambda_k-{\pi i\over12})}}.
\eeq

These equations have been solved to find the order one excitation
energies \cite{Al1}.  The results are expressed in terms of three
single particle excitation energies:
\subequations
\beq
e_{2s}(P)=3\{\sqrt2 \cos ({{|P|}\over 2}-{{3 \pi}\over4})+1\},
\eeq
\beq
 e_{-2s}(P)=3\{\sqrt2 \cos({{|P|}\over2}-{{\pi}\over4})-1\},
\eeq
\beq
e_{ns}(P)=3 \sin({{|P|}\over2}).
\eeq
\endsubequations
For $P\sim 0$, all three excitations are of the form (1.5) with
\beq
v=3/2.
\eeq
Here and in the remainder of the paper we take $M$ to be even.
\begin{enumerate}
\item For $Q=0$ the energies and momenta are of the form (1.2) and (1.3)
\beq
E(\{P_j^{2s}\},\{P_j^{-2s}\},\{P_j^{ns}\})-E_{GS}=\sum_{\alpha=2s,-2s,ns}
\sum_{j=1}^{m_{\alpha}}e_{\alpha}(P_j)
\eeq
and
\beq
P=P^0+\sum_{\alpha=2s,-2s,ns}\sum_{j=1}^{m_{\alpha}}P_j^{\alpha},
\eeq
where
\beq
P^0= P_{GS}={M\over2}\pi\quad {\rm mod}2\pi,
\eeq
\beq
m_{2s}+m_{-2s}\quad {\rm is \quad even},
\eeq
$P_j^{2s}$, $P_j^{-2s}$, and $P_j^{ns}$ obey the fermi exclusion rule
(1.4) and they lie in the ranges
\subequations
\beq 0\leq P_j^{2s}\leq3\pi,\eeq
\beq 0\leq P_j^{-2s}\leq\pi,\eeq
\beq 0\leq P_j^{ns}\leq 2\pi.\eeq
\endsubequations
We also note that the C parity of the ground state is
\beq
C_{GS}=(-1)^{M/2},
\eeq
and the C parity of an arbitrary state is
\beq
C/C_{GS}=(-1)^{m_{ns}+m_{-2s}+(m_{2s}+m_{-2s})/2}.
 \eeq
\item For $Q=\pm1$  we must consider $m_{2s}+m_{-2s}$ to
be both even and odd.  When $m_{2s}+m_{-2s}$ is even there are two
spectra of the form (2.14) and (2.15).  In one $P^0=P_{GS} $ and in
the other $P^0=P_{GS}+\pi$. In both cases the $P_j^{\alpha}$ obey
(2.18).  When $m_{2s}+m_{-2s}$ is odd, there are again two spectra of
the form (2.14) and (2.15).  In each case $P^0=P_{GS}$. In one case
$P_j^{\alpha}$ satisfies (2.18), while in the other case
$-P_j^{\alpha}$ satisfies (2.18).

\end{enumerate}

The conformal field theory predictions for the model can be obtained
by noting that the 3-state Potts model is the critical $D_4$ model in
the classification of Pasquier \cite{Pas}. The central charge and the
conformal dimensions of the primary fields are thus obtained by
specializing the finite size computations of the $A_{N+1}$ model of
Kl{\"u}mper and Pearce \cite{KP2} to the case $N=4$ and using an orbifold
construction \cite{FG} to find the results for $D_4$. The
general result for $A_N+1$ at the boundary of the I/II regime is that the
central charge is
\beq
c={2(N-1)\over N+2},
\eeq
and the conformal dimensions are
\beq
h_m^l={\ell(\ell+2)\over 4(N+2)}-{m^2\over 4N}\quad {\rm for}\quad
|m|\leq \ell
\eeq
which are the same as those of the $Z_N$ parafermion conformal field
theory of Zamolodchikov and Fateev~\cite{zf}.
Using the symmetry $h_m^{\ell}=h_{N-m}^{N-\ell}$ we find for $N=4$:
\beq
c=1,
\eeq
and
\beq
h_0^0=0,\quad h_2^0={3\over 4},\quad h_4^0=1,\quad h_0^2={1\over 3},
\quad h_2^2={1\over 12},
\eeq
where the first three conformal dimensions occur in $Q=0$ and the last
two in $Q=\pm1$.  Moreover the modular invariant partition function is
that of the $D_4$ parafermion model \cite{GQ} (1.10) where the
branching functions $b_m^\ell$ can be obtained by specializing to
$N=4$ the Hecke indefinite form of Kac-Peterson
\cite{Kac1}
\begin{eqnarray}
b_m^\ell &=& Q(q)^{-2}
q^{{l(l+2)\over 4(N+2)}-\frac{m^2}{4N}-\frac{c}{24}}\nonumber \\
&\times&\Bigg[\Bigg(\sum_{s\geq0}\sum_{n\geq0}-\sum_{s<0}\sum_{n<0}\Bigg)
(-1)^s q^{s(s+1)/2+(l+1)n+(l+m)s/2+(N+2)(n+s)n}\nonumber \\
&+& \Bigg( \sum_{s>0}\sum_{n\geq0}-\sum_{s\leq0}\sum_{n<0}\Bigg)
(-1)^s q^{s(s+1)/2+(l+1)n+(l-m)s/2+(N+2)(n+s)n}\Bigg]
\end{eqnarray}
for $|m|\leq \ell$, and using the symmetries
\beq
b_m^{\ell}=b_{-m}^{\ell}=b_{m+2N}^{\ell}=b_{N-m}^{N-\ell}
\eeq
otherwise.  An alternative form for $b_m^{\ell}$ is given in \cite{DQ}
but for our purposes the simplest form is the specialization which
only occurs for $N=4$
\cite{KW}
\beq
b_0^0+b_4^0=f_{3,0}/\eta, \quad b_0^0-b_4^0=g_{1,0}/\eta, \quad
b_2^0=f_{3,3}/2\eta, \quad b_0^2=f_{3,2}/\eta, \quad  b_2^2=f_{3,1}/\eta,
\eeq
where
\beq \eta=q^{{1\over24}}Q(q), \eeq
and
\beq f_{a,b}=\sum_{n=-\infty}^{\infty}q^{a(n+{b\over 2a})^2}, \quad
 g_{a,b}=\sum_{n=-\infty}^{\infty}(-1)^n q^{a(n+{b\over 2a})^2}.
\eeq
This form has a simple origin in the Gaussian model with $r=\sqrt
{3/2}$, which we give in appendix A. For comparison with the
expansions of subsequent sections, we list the first few terms of
(2.27) as
\subequations
\beq
q^{1/24}b_0^0=1+q^2+2q^3+4q^4+5q^5+9q^6+12q^7+19q^8+25q^9+37q^{10}+\cdots) \eeq
\beq
q^{1/24}b_2^0=q^{3/4}(1+q+2q^2+3q^3+5q^4+7q^5+12q^6+16q^7+24q^8+33q^9+47 q^{10}
+\cdots ) \eeq
\beq q^{1/24}b_4^0=q(1+q+3q^2+3q^3+6q^4+8q^5+13q^6+17q^7+27q^8+ 35 q^9+51
q^{10}+\cdots ) \eeq
\beq
q^{1/24}b_0^2=q^{1/3}(1+2q+3q^2+5q^3+8q^4+13q^5+19q^6+28q^7+41q^8+58q^9+
81q^{10}+\cdots)
\eeq
\beq q^{1/24}b_2^2=q^{1/12}(1+q+3q^2+4q^3+8q^4+11q^5+18q^6+25q^7+38q^8+52q^9+
76q^{10}+\cdots)\eeq
\endsubequations

We finally note that there is an alternative way to obtain these
conformal field theory predictions which utilizes $W_4$
algebra~\cite{fateev} and is related to the GKO construction
${(A_3^{(1)})_1\times (A_3^{(1)})_1\over (A_3^{(1)})_2}$. The
branching functions have been computed from this construction in terms
of three dimensional sums~\cite{jimbo,cnr,fy2} which, for later
comparison, we give in appendix B.

\section{Branching Functions  for Q=0}
\setcounter{equation}{0}
The partition function for the Hamiltonian (2.1) is, by definition,
\beq
Z=\sum_{n}e^{-E_n/k_B T}=e^{-E_{GS}/k_B T}\sum e^{-(E_{n}-E_{GS})/k_B T},
\eeq
where for $M\rightarrow\infty$,
\beq
E_{GS}=Me_0-{\pi cv\over6M} + O(\frac{1}{M^2})
\eeq
and from~\cite{KP2} $c=1$. To obtain the relation with the modular
invariant partition function of conformal field theory we must
evaluate (3.1) in the limit $M\rightarrow\infty$, $T\rightarrow 0$
with $MT$ fixed. We intend to carry out this evaluation by making use
of the quasiparticle energy spectrum (2.14).

There are, however, two questions that must be addressed before we can
do this. The first is that in order for (2.14) to specify the energy
levels, the momenta $P_j^{\alpha}$ must be discretely specified on the
finite lattice. The second is that the evaluation leading to (2.14) is
only correct to order one as $M\rightarrow \infty$ and hence in order
to agree with~\cite{KP2} it may be necessary to add some term of the
order of $1/M$ which is independent of $P_j^{\alpha}$ but which in
general will depend on $m_{2s}$, $m_{-2s}$ and $m_{ns}$. These
considerations are different for $Q=0$ and $Q=\pm 1$.

We consider in this section $Q=0$. We find from the previous
study~\cite{Al2} of the competeness of the solutions of (2.1) that for
given $m_{2s}$, $m_{-2s}$, and $m_{ns}$
\subequations
\beq
P_j^{2s}\quad {\rm takes}\quad
{3M\over2}-m_{ns}-{m_{2s}+m_{-2s}\over2}\quad {\rm values},
\eeq
\beq
P_j^{-2s}\quad {\rm takes}\quad
{M\over2}-m_{ns}-{m_{2s}+m_{-2s}\over2}\quad {\rm values},
\eeq
and
\beq
P_j^{ns}\quad {\rm takes} \quad M-m_{ns}-m_{2s}-m_{-2s}\quad {\rm
values}. \eeq
\endsubequations
This will be the case if $P_j^{\alpha}$ satisfies
\subequations
\beq
{\pi\over M}(m_{ns}+{m_{2s}+m_{-2s}\over 2}+1)\leq P_j^{2s}\leq
3\pi-{\pi\over M}(m_{ns}+{m_{2s}+m_{-2s}\over 2}+1),
\eeq
\beq
{\pi\over M}(m_{ns}+{m_{2s}+m_{-2s}\over 2}+1)\leq P_j^{-2s}\leq
\pi-{\pi\over M}(m_{ns}+{m_{2s}+m_{-2s}\over 2}+1),
 \eeq
\beq
{\pi\over M}(m_{ns}+m_{2s}+m_{-2s}+1)\leq P_j^{ns}\leq
2\pi-{\pi\over M}(m_{ns}+m_{2s}+m_{-2s}+1),
 \eeq
\endsubequations
where the spacing between allowed values for $P_j^\alpha$ is
$2\pi/M$, and $m_{2s}+m_{-2s}$ is even.  It may be verified that this
choice of $P_j^{\alpha}$ exactly reproduces the correct number of
momenta of table~4 of~\cite{Al2} for each allowed set of $m_{2s}$,
$m_{-2s}$ and $m_{ns}$.

Since $M\rightarrow\infty$ and $T\rightarrow 0$ with $MT$ fixed, only
those values of $P_j^{\alpha}$ where $e_{\alpha}(P)$ is small of the
order $1/M$ contributes to (3.1). This occurs for
\subequations
\beq
P_j^{\alpha}\sim 0\quad {\rm for}\quad \alpha=2s,-2s, ns,
\eeq
\beq
P_j^{-2s}\sim \pi,\quad P_j^{ns}\sim 2\pi\quad,P_j^{2s}\sim 3\pi,
\eeq
\endsubequations
where we may linearize $e_{\alpha}(P)$ near the endpoints (3.5) as
\subequations
\beq
e_{\alpha}(P)\sim vP^{\alpha}\quad {\rm for}\quad P^{\alpha}\quad{\rm near}
\quad {\rm zero},
\eeq
\beq
 e_{-2s}(P)\sim v(\pi-P^{-2s})\quad {\rm for}\quad P^{-2s}\quad
{\rm near}\quad \pi ,
\eeq
\beq
e_{ns}(P)\sim v(2\pi-P^{ns})\quad{\rm for}\quad P^{ns}\quad {\rm
near}\quad 2\pi ,
\eeq
and
\beq
 e_{2s}(P)\sim v(3\pi-P^{2s})\quad {\rm for}\quad P^{2s}
\quad {\rm near}\quad 3\pi.
\eeq
\endsubequations

Thus we let $m_{\alpha}^{\ell}$ be the number of $P_j^{\alpha}$ near
zero and $m_{\alpha}^r$ be the number of $P_j^{\alpha}$ near the end
points (3.5b).  We note that
\beq
m_{\alpha}^{\ell}+m_{\alpha}^r=m_{\alpha}.
\eeq
We also note that if
\beq
m_{2s}^r+m_{-2s}^r\quad {\rm is} \quad {\rm odd}
\eeq
then from (2.15) the total momentum of the state is macroscopically
shifted $\pi$ from the ground state value $P_{GS}$. These states are
expected to make oscillatory contributions to the correlation
functions.

Consider first the case where all $m_{\alpha}^r=0$ (which by symmetry
is identical to the case $m_{\alpha}^{\ell}=0$) and evaluate the
partition function (3.1) using (2.14), (3.4), and (3.6) in the case
$C/C_{GS}=1$ where by (2.20)
\beq
m_{ns}^{\ell}+m_{-2s}^{\ell}+(m_{2s}^{\ell}+m_{-2s}^{\ell})/2\quad {\rm is
\quad even}.
\eeq
We present in table~1 the terms from this construction up through
order $q^8$ where we see that they agree with the corresponding terms
from the branching function $q^{1/24}b_0^0$ of (2.27). This equality
has been verified to order $q^{200}$ and thus we conclude that this
construction correctly gives the branching function $b_0^0$.

To obtain a formula for $b_0^0$ from the construction, let
$P_{d}(m,n)$ denote the number of distinct ways that the integer $n$
can be additively partitioned into $m$ distinct parts. Then, modifying the
usual construction of a free fermi partition function in terms of
$P_d(m,n)$ to account for the momentum exclusion rule (3.4), we find
\begin{eqnarray}
q^{1/24}b_0^0&=&
\sum_{m_{ns},m_{2s},m_{-2s}=0}^\infty
\sum_{n_{ns},n_{2s},n_{-2s}=0}^\infty
P_d(m_{ns},n_{ns})P_d(m_{2s},n_{2s})P_d(m_{-2s},n_{-2s})
q^{n_{ns}+n_{2s}+n_{-2s}}\nonumber \\ & &\quad \quad \times
q^{{m_{ns}\over 2}(m_{ns}+m_{2s}+m_{-2s}-1)} q^{{(m_{2s}+m_{-2s})\over
2}(m_{ns}+\frac{m_{2s}+m_{-2s}}{2}-1)}
\end{eqnarray}
where $m_{2s}+m_{-2s}$ and $m_{ns}+m_{-2s}+(m_{2s}+m_{-2s})/2$ are
even and $P_d(0,0)=1$ by definition.  The sums over $n_{\alpha}$ are
evaluated using
\beq
\sum_{n=0}^{\infty}P_d(m,n)q^n=\frac{q^{m(m+1)/2}}{(q)_m},
\eeq
where we use the standard notation
\beq
(q)_m=\prod_{j=1}^{m}(1-q^j),
\eeq
and $(q)_0=1$ by definition. Thus we find when $m_{2s}+m_{-2s}$ is
even and (3.9) holds
\begin{eqnarray}
q^{1/24}b_0^0&=&\sum_{m_{ns}=0}\sum_{m_{2s}=0}\sum_{m_{-2s}=0}
{q^ {m_{ns}(m_{ns}+1)/2}\over (q)_{m_{ns}}}
{q^{m_{2s}(m_{2s}+1)/2}\over(q)_{m_{2s}}}
{q^{m_{-2s}(m_{-2s}+1)/2}\over(q)_{m_{-2s}}}\nonumber\\
& &\quad \quad q^{{m_{ns}\over 2}(m_{ns}+m_{2s}+m_{-2s}-1)}
q^{{(m_{2s}+m_{-2s})\over 2}(m_{ns}+(m_{2s}+m_{-2s})/2-1)}.
\end{eqnarray}

We may now extend these considerations to the general case where both
some $m_{\alpha}^r\neq 0$ and some $m_{\alpha}^{\ell}\neq 0$. In this
general case we note from the work of~\cite{KP1,KP2} that the
contribution to the energy from regions where (3.5a) holds and the
region where (3.5b) holds are independent. Combining the above
considerations we have in general the expression for the low lying
energy levels in the $M\rightarrow \infty$ limit:
\beq
E_{ex}-E_{GS}=\sum_{\alpha=2s,-2s,ns}\left\{\sum_{j=1}^{m_{\alpha}^{\ell}}
e_{\alpha}(P_j^{{\ell.\alpha}})+\sum_{j=1}^{m_{\alpha}^r}
e_{\alpha}(P_j^{r,\alpha})\right\},
\eeq
where we define $P_j^{\ell ,\alpha}$ and $P_j^{r,\alpha}$ to satisfy
\subequations
\beq
{\pi\over M}(m_{ns}^{\ell}+{m_{2s}^{\ell}+m_{-2s}^{\ell}\over2}+1)
\leq P_j^{\ell,2s},P_j^{\ell,-2s},
\eeq
\beq
{\pi\over M}(m_{ns}^{\ell}+m_{2s}^{\ell}+m_{-2s}^{\ell}+1)
\leq P_j^{\ell,ns},
\eeq
\endsubequations
and
\subequations
\beq
{\pi\over M}(m_{ns}^r+{m_{2s}^r+m_{-2s}^r\over 2}+1)\leq P_j^{r,2s},
P_j^{r,-2s},
\eeq
\beq
{\pi\over M}(m_{ns}^r+m_{2s}^r+m_{-2s}^r+1)\leq P_j^{r,ns},
\eeq
\endsubequations
where again, the spacing between allowed values for $P_j^{l,r}$ is
$2\pi/M$ and $e_{\alpha}(P)=v P $ with $v$ given by (2.13).

In table~2 we consider the cases $m_{2s}^r=1$, $m_{-2s}^r=m_{ns}^r=0$
and $m_{-2s}^r=1$, $m_{2s}^r=m_{ns}^r=0$ and compute the contribution
to made to $Z$ to order $q^{31/4}$ of the terms in (3.14) that involve
only $P_j^{\ell , \alpha}$.  From (2.17) we see that
\beq
m_{2s}^{\ell}+m_{-2s}^{\ell}\quad {\rm is \quad odd}.
\eeq
Further, the interchange $m_{2s}^{\ell}\leftrightarrow m_{-2s}^{\ell}$
leaves (3.14) invariant and from (2.20) gives $C \leftrightarrow -C$.
Thus we need only consider $m_{2s}^{\ell}< m_{-2s}^{\ell}$ and find
that this construction agrees with $q^{1/24}b_2^0$ of (2.30b).

In table~3 we consider the case $m_{ns}^r=1$, $m_{2s}^r=m_{-2s}^r=0$
and compute to order $q^8$ the contribution made to $Z$ in the channel
$C/C_{GS}=1$ of the terms in (3.14) that involve only $P_j^{\ell,
\alpha}$. From (2.7) and (2.20) we find that
\beq
m_{2s}^{\ell}+m_{-2s}^{\ell}\quad {\rm even \quad
and}\quad m_{ns}^{\ell}+m_{-2s}^{\ell}+(m_{2s}^{\ell}
+m_{-2s}^{\ell})/2\quad {\rm odd}.
\eeq
We find that this  agrees with $q^{1/24}b_4^0$ of (2.30c).

These above two equalities have been verified to order $q^{200}$.

{}From these constructions we can find expressions for $b_2^0$ and
$b_4^0$ as we did above for $b_0^0$.  We thus find that
$q^{1/24}b_{\alpha}^0$ is given by (3.13) for $\alpha =0,2,4$ where
\begin{eqnarray}
{\rm for}\quad &b_0^0& \quad m_{2s}+m_{-2s}\quad {\rm is \quad
even \quad and}\quad m_{ns}+m_{-2s}+\frac{m_{2s}+m_{-2s}}{2}\quad {\rm is \quad
even},\nonumber\\
{\rm for}\quad &b_2^0& \quad m_{2s}+m_{-2s}\quad {\rm is\quad odd\quad and}
\quad m_{2s}< m_{-2s},
\nonumber\\
{\rm for}\quad&b_4^0& \quad m_{2s}+m_{-2s}\quad
{\rm is \quad even\quad and }\quad
m_{ns}+m_{-2s}+\frac{m_{2s}+m_{-2s}}{2}\quad {\rm is \quad odd}.
\end{eqnarray}

We may now finally construct the complete $Q=0$ contribution to $Z$ by
using (3.14)-(3.16) in (3.1) and summing over all $m_{\alpha}^r$ and
$m_{\alpha}^{\ell}$ subject only to the restriction (2.17) written in
the form
\beq
m_{2s}^r+m_{-2s}^r+m_{2s}^{\ell}+m_{-2s}^{\ell}\quad {\rm even}.
\eeq
(We note that there is no restriction corresponding to (2.20) because
both channels $C=\pm1$ are considered in the sum.) It is easy then to
see that the result consists of all the terms in (1.10) which involve
$b_0^0$ ,$b_2^0$ and $b_4^0$ where we note: 1) that the factor of 4 in
front of $b_2^0(q)b_2^0({\bar q})$ arises because of the symmetry
under $m_{2s}^r\leftrightarrow m_{-2s}^r$ and
$m_{2s}^{\ell}\leftrightarrow m_{-2s}^{\ell}$, and 2) terms like
$b_0^0(q)b_2^0({\bar q})$ and $b_4^0(q)b_2^0({\bar q})$ are excluded
by (3.20).

\section{Branching Functions for Q=1}
\setcounter{equation}{0}
The channel $Q={\pm1}$ is more complicated that the channel $Q=0$
because as seen in section~2 the spectrum of excitations has 4
separate contributions. In~\cite{Al2} these contributions are
distinguished by the number $m_{++}$ of $(++)$ pairs of roots, and the
number $m_{-+}$ of $(-+)$ pairs of roots where there is the sum rule:
\beq
m_{2s}+2m_{ns}+3m_{-2s}+m_{-+}+m_{++}=M-1.
\eeq
We found that the three cases occurred of
\beq
m_{-+}-m_{++}=1,0,-1,
\eeq
and when $m_{-+}=m_{++}$ the spectrum is two fold degenerate. We will
thus extend the considerations of section~3 by considering these three
cases separately.

\subsection{$\bf m_{-+}-m_{++}=-1$}
In this sector the total momentum is given by (2.15) with
$P^0=P_{GS}+\pi$ and there are $M-1$ single particle states with
$m_{ns}=1$.

We find for all three cases (4.2) from the previous study of
completeness~\cite{Al2} that
\subequations
\beq
P_j^{2s}\quad{\rm takes}\quad M-1+m_{++}+m_{-2s} \quad {\rm values},
\eeq
\beq
P_j^{-2s}\quad {\rm takes} \quad m_{-2s}+m_{++}-1 \quad {\rm values},
\eeq
and
\beq
P_j^{ns}\quad {\rm takes}\quad m_{ns}+2m_{++}+2m_{-2s}\quad {\rm values}.
\eeq
\endsubequations

In this present case we use $m_{-+}=m_{++}-1$ in (4.1) to write
\beq
2m_{++}=M-m_{2s}-2m_{ns}-3m_{-2s},
\eeq
and thus (4.3) reduces to
\subequations
\beq
{3M\over2}-1-m_{ns}-{m_{2s}+m_{-2s}\over2 }\quad {\rm values \quad
for \quad}P_j^{2s},
\eeq
\beq
{M\over2}-1-m_{ns}-{m_{2s}+m_{-2s}\over2} \quad {\rm values \quad
for \quad}P_j^{-2s},
\eeq
and
\beq
M-m_{ns}-m_{2s}-m_{-2s} \quad {\rm values \quad for }\quad
P_j^{ns},
\eeq
\endsubequations
where $m_{2s}+m_{-2s}$ is even.This will be the case if $P_j^{\alpha}$
satisfies  :
\subequations
\beq
{\pi\over M}(m_{ns}+{m_{2s}+m_{-2s}\over 2}+2)\leq P_j^{2s}\leq
3\pi-{\pi\over M}(m_{ns}+{m_{2s}+m_{-2s}\over 2}+2),
\eeq
\beq
{\pi\over M}(m_{ns}+{m_{2s}+m_{-2s}\over 2}+2)\leq P_j^{-2s}\leq
\pi-{\pi\over M}(m_{ns}+{m_{2s}+m_{-2s}\over 2} +2),
\eeq
and
\beq
{\pi\over M}(m_{ns}+m_{2s}+m_{-2s}+1)\leq P_j^{ns}\leq 2\pi
-{\pi\over 2}(m_{ns}+m_{2s}+m_{-2s}+1),
\eeq
\endsubequations
where
\beq
P_{j+1}^{\alpha}-P_j^{\alpha}=2\pi/M.
\eeq
It may be verified that this choice of $P_j^{\alpha}$ exactly
reproduces the momenta of table~6 of~\cite{Al2} for each allowed set
of $m_{2s}$, $m_{-2s}$ and $m_{ns}$.  Following the procedure of
section~3 we compute in table~4 the contribution these states will
make to the partition function where we use the linearized energies
(3.6) and keep all $P_{j}^{\alpha}$ near zero.

\subsection{$\bf m_{-+}-m_{++}=1$}

In this case $P^0=P_{GS}$ and there are $M-3$ single particle states
with $m_{ns}=1$. Furthermore we find from (4.1) that
\beq
2m_{++}=M-2-m_{2s}-2m_{ns}-3m_{-2s},
\eeq
and thus there are
\subequations
\beq
{3M\over 2} -2-m_{ns}-{m_{2s}+m_{-2s}\over 2} \quad {\rm values
\quad for\quad}P_j^{2s},
\eeq
\beq
{M\over 2} -2-m_{ns}-{m_{2s}+m_{-2s}\over 2} \quad {\rm values
\quad \quad } P_j^{-2s},
\eeq
and
\beq
M-2-m_{ns}-m_{2s}-m_{-2s} \quad {\rm values \quad for \quad}P_j^{ns},
\eeq
\endsubequations
where again $m_{2s}+m_{-2s}$ is even. This will be satisfied if
$P_j^{\alpha}$ satisfies
\subequations
\beq
{\pi\over M}(m_{ns}+{m_{2s}+m_{-2s}\over 2}+3)\leq P_j^{2s}\leq 3\pi -
{\pi \over M}(m_{ns}+{m_{2s}+m_{-2s}\over 2}+3),
\eeq
\beq
{\pi\over M}(m_{ns}+{m_{2s}+m_{-2s}\over 2}+3)\leq P_j^{-2s}\leq
\pi -{\pi\over M}(m_{ns}+{m_{2s}+m_{-2s}\over 2}+3),
\eeq
\beq
{\pi\over M}(m_{ns}+m_{2s}+m_{-2s}+3)\leq P_j^{ns}\leq 2\pi
-{\pi\over M}(m_{ns}+m_{2s}+m_{-2s}+3),
\eeq
\endsubequations
where (4.7) holds. Again it may be verified that this choice of
$P_j^{\alpha}$ exactly reproduces the momenta of table~6 of~\cite{Al2}
for the allowed values of $m_{2s}$, $m_{-2s}$ and $m_{ns}$. In table~5
we compute the contribution these states make to the partition
function where the linearized energies (3.6) are used and all momenta
are kept near zero.

\subsection{$\bf m_{-+}=m_{++}$}
In this case $P^0=P_{GS}$, there are $3M-4$ single particle states
with $m_{2s}=1$ and $M-4$ single particle states with $m_{-2s}=1$.
Furthermore
\beq
2m_{++}=M-1-m_{2s}-2m_{ns}-3m_{-2s},
\eeq
where now $m_{2s}+m_{-2s}$ is odd. In this case there is a double
degeneracy and we find from~\cite{Al2} that there are
\subequations
\beq
2\times ({3M\over 2}-1-m_{ns}-{m_{2s}+m_{-2s}+1\over 2}) \quad
{\rm values \quad for \quad}P_j^{2s},
\eeq
\beq
2\times ({M\over 2}-1-m_{ns}-{m_{2s}+m_{-2s}+1\over 2})\quad {\rm
values \quad for\quad}P_j^{-2s},
\eeq
and
\beq
2\times (M-1-m_{ns}-m_{2s}-m_{-2s})\quad {\rm values \quad for
\quad}P_j^{ns}.
\eeq
\endsubequations

Now in order to get a formula for the momentum which respects (4.12)
we must consider two subcases. Either $P_j^{\alpha}$ satisfy
\subequations
\beq
{\pi\over M}(m_{ns}+{m_{2s}+m_{-2s}+1\over 2}+1)\leq P_j^{2s}\leq
3\pi-{\pi\over M}(m_{ns}+{m_{2s}+m_{-2s}+1\over 2} +3),
\eeq
\beq
 {\pi\over M}(m_{ns}+{m_{2s}+m_{-2s}+1\over 2}+1)\leq P_j^{-2s}\leq
\pi-{\pi\over M}(m_{ns}+{m_{2s}+m_{-2s}+1\over 2}+3),
\eeq
\beq
 {\pi\over M}(m_{ns}+m_{2s}+m_{-2s}+1)\leq P_j^{ns}\leq
2\pi-{\pi\over M}(m_{ns}+m_{2s}+m_{-2s}+3),
\eeq
\endsubequations
or $P_j^{\alpha}$ satisfy
\subequations
\beq
-3\pi+{\pi\over M}(m_{ns}+{m_{2s}+m_{-2s}+1\over 2}+3)\leq
P_j^{2s}\leq -{\pi\over M}(m_{ns}+{m_{2s}+m_{-2s}+1\over 2}+1),
\eeq
\beq
-\pi +{\pi\over M}(m_{ns}+{m_{2s}+m_{-2s}+1\over 2}+3)\leq
P_j^{-2s}\leq -{\pi\over M}(m_{ns}+{m_{2s}+m_{-2s}+1\over 2}+1),
\eeq
\beq
 -2 \pi+{\pi\over M}(m_{ns}+m_{2s}+m_{-2s}+3)\leq P_j^{ns}\leq
-{\pi\over M}(m_{ns}+m_{2s}+m_{-2s}+1).
\eeq
\endsubequations

However, now instead of the total momentum $P$ being given in terms of
$P_j^{\alpha}$  by (2.15) we must introduce a shift   of order $1/M$
(which is permissible because (2.15) is only derived to order one) and
write that when (4.15) holds
\subequations
\beq
 P=P^0+\sum_{\alpha=2s,-2s,ns}\sum_{j=1}^{m_{\alpha}}P_j^{\alpha}
+{\pi\over M}({m_{2s}+m_{-2s}-1\over2}),
\eeq
and when (4.16) holds
\beq
P=P^0+\sum_{\alpha=2s,-2s,ns}\sum_{j=1}^{m_{\alpha}}P_j^{\alpha}
-{\pi\over M}({m_{2s}+m_{-2s}-1\over 2}).
\eeq
\endsubequations
It may be verified that the momenta computed from these rules agrees
with the momenta of table~6 of~\cite{Al2} for the allowed values of
$m_{2s}$, $m_{-2s}$ and $m_{ns}$.

Corresponding to this momentum shift there is an energy shift as
well. Thus in table~6 we compute the contribution to the partition
function of those states obtained from (4.13) with $P_j^{\alpha}$ near
zero using the linearized energies (3.6) and subtracting the shift
${\pi\over M}({m_{2s}+m_{-2s}-1\over 2})$. We note that the
macroscopic momentum of these states are near $P_{GS}$. In table~7 we
compute the contribution to the partition function from those states
obtained from (4.14) with $P_j^{2s}$ near $-3\pi$, $P_j^{-2s}$ near
$-\pi$ and $P_j^{ns}$ near $-2\pi$ where we add a shift ${\pi\over
M}({m_{2s}+m_{-2s}-1\over 2})$ to the energy.  The macroscopic
momentum of these states is $P_{GS}+\pi$. In both cases, due to the
symmetry under $2s\leftrightarrow -2s$, only the states with
$m_{2s}<m_{-2s}$ are shown.

\subsection{Branching functions}
We may now obtain the formulas for the two branching functions of
$Q=1$, namely $b_2^0$ and $b_2^2$ by combining together the results
of the three preceding subsections with the same macroscopic
momentum.

Consider first the case where the macroscopic momentum is $P_{GS}$.
This is obtained from the $m_{-+}-m_{++}=1$ states of table~5 and the
$m_{-+}=m_{++}$ states of table~6. In table~8 we compute the sum of
these two contributions and see that it is identical with the
corresponding terms in $q^{-1/3}q^{1/24}b_0^2$ of (2.30d).

The other case has the macroscopic momentum of $P_{GS}+\pi$ .This is
obtained from the $m_{-+}-m_{++}=-1$ states of table~4 and the
$m_{-+}=m_{++}$ states of table~7. In table~9 we compute the sum of
these two contributions and see that it is identical with the
corresponding terms in $q^{-1/12}q^{1/24}b_2^2$ of (2.30e).

As in $Q=0$ these identities have been verified to order $q^{200}$.

We may now use the above construction to obtain formulas for the
branching functions $b_0^2$ and $b_2^2$ by following exactly the same
procedure used in section~3. Thus we find
\newpage
\subequations
\begin{eqnarray}
q^{-1/3}q^{1/24} b_0^2
&=&\sum_{m_{ns}=0}^\infty\sum_{m_{2s}=0}^\infty
\sum_{\stackrel{m_{-2s}=0}{m_{2s}+m_{-2s}\,{\rm even}}}^\infty
{q^{m_{ns}(m_{ns}+1)\over 2}\over (q)_{m_{ns}}}
{q^{m_{2s}(m_{2s}+1)\over 2}\over (q)_{m_{2s}}}
{q^{m_{-2s}(m_{-2s}+1)\over 2}\over (q)_{m_{-2s}}}\nonumber\\
& &\quad\quad \times q^{{m_{ns}\over
2}(m_{ns}+m_{2s}+m_{-2s}+1)}q^{{(m_{2s}+m_{-2s})\over 2}
(m_{ns}+{m_{2s}+m_{-2s}\over 2}+1)}\nonumber\\
&+&\sum_{m_{ns}=0}^\infty\sum_{m_{2s}=0}^\infty
\sum_{\stackrel{m_{-2s}=0}{m_{2s}+m_{-2s}\,{\rm odd}}}^\infty
{q^{m_{ns}(m_{ns}+1)\over 2}\over (q)_{m_{ns}}}
{q^{m_{2s}(m_{2s}+1)\over 2}\over (q)_{m_{2s}}}
{q^{m_{-2s}(m_{-2s}+1)\over 2}\over (q)_{m_{-2s}}}\nonumber\\
& &\quad\quad \times q^{{m_{ns}\over 2}(m_{ns}+m_{2s}+m_{-2s}-1)}
q^{{(m_{2s}+m_{-2s})\over 2}(m_{ns}+{m_{2s}+m_{-2s}\over 2})}q^{-1/4}
\end{eqnarray}
and
\begin{eqnarray}
q^{-1/12}q^{1/24}b_2^2
&=&\sum_{m_{ns}=0}^\infty\sum_{m_{2s}=0}^\infty
\sum_{\stackrel{m_{-2s}=0}{m_{2s}+m_{-2s}\,{\rm even}}}^\infty
{q^{m_{ns}(m_{ns}+1)\over 2}\over (q)_{m_{ns}}}
{q^{m_{2s}(m_{2s}+1)\over 2}\over (q)_{m_{2s}}}
{q^{m_{-2s}(m_{-2s}+1)\over 2}\over (q)_{m_{-2s}}}\nonumber\\
& &\quad\quad \times q^{{m_{ns}\over
2}(m_{ns}+m_{2s}+m_{-2s}-1)}q^{{(m_{2s}+m_{-2s})\over 2}
(m_{ns}+{m_{2s}+m_{-2s}\over 2})}\nonumber\\
&+&\sum_{m_{ns}=0}^\infty\sum_{m_{2s}=0}^\infty
\sum_{\stackrel{m_{-2s}=0}{m_{2s}+m_{-2s}\,{\rm odd}}}^\infty
{q^{m_{ns}(m_{ns}+1)\over 2}\over (q)_{m_{ns}}}
{q^{m_{2s}(m_{2s}+1)\over 2}\over (q)_{m_{2s}}}
{q^{m_{-2s}(m_{-2s}+1)\over 2}\over (q)_{m_{-2s}}}\nonumber\\
& &\quad\quad\times q^{{m_{ns}\over 2}(m_{ns}+m_{2s}+m_{-2s}+1)}
q^{{(m_{2s}+m_{-2s})\over 2}(m_{ns}+{m_{2s}+m_{-2s}\over 2}+1)}q^{1/4}.
\end{eqnarray}
\endsubequations

\subsection{Partition function}
Finally we need to compute the complete partition function in the
$Q=1$ channel. Now in each of the four cases considered above there
are both left and right excitations $m_{\alpha}^{r}$ and
$m_{\alpha}^{\ell}$. Denote the four sums obtained above for all
$m_{\alpha}^{r}=0$ by $S^{-}$ for $m_{-+}-m_{++}=-1$,by $S^{+}$ for
$m_{-+}-m_{++}=1$, by $S^0$ for $m_{-+}=m_{++}$ with $\Delta P =0$,
and by $S^{\pi}$ for $m_{-+}=m_{++}$ with $\Delta P =\pi$. To consider
the general case with both $m_{\alpha}^r$ and $m_{\alpha}^{\ell}$ both
nonzero, we follow the procedure of Section~3 and consider momentum
restrictions for the right and left movers separately.

Consider first $m_{-+}-m_{++}=-1$. Then because of the restriction
that $m_{2s}+m_{-2s}=m_{2s}^r+m_{-2s}^r+m_{-2s}^{\ell}+m_{-2s}^{\ell}$
must be even we see that $m_{2s}^r+m_{-2s}^r$ and
$m_{2s}^{\ell}+m_{-2s}^{\ell}$ must be even or odd together. If both
terms are even the contribution made to the partition function is
$S^{-}(q)S^{-}({\bar q})$ and if both terms are odd and the momenta
are shifted properly the contribution is $S^{\pi}(q)S^{\pi}({\bar
q})$.

The contribution from $m_{-+}-m_{++}=1$ is similar. Again
$m_{2s}^r+m_{-2s}^r$ and $m_{2s}^{\ell}+m_{-2s}^{\ell}$ must be even
or odd together. The even terms contribute $S^{+}(q)S^{+}({\bar q})$
and the odd terms contribute $S^{0}(q)S^{0}({\bar q})$.

Finally there are the terms from $m_{++}=m_{-+}$. Here
$m_{2s}^r+m_{-2s}^r+m_{2s}^{\ell}+m_{-2s}^{\ell}$ is odd, thus
$m_{2s}^r+m_{-2s}^r$ is even (odd) and $m_{2s}^{\ell}+m_{-2s}^{\ell}$
is odd (even). The ends where $m_{2s}^{r,\ell}+m_{-2s}^{r,\ell}$ are
odd produce the terms $S^{0}$ and $S^{\pi}$ and if the momentum shift
is properly accounted for the ends with
$m_{2s}^{r,\ell}+m_{-2s}^{r,\ell}$ even produce $S^+$ and $S^-$. Thus
the sector $m_{++}=m_{-+}$ produces the cross terms like
$S^0(q)S^+({\bar q})$ and $S^{\pi}(q)S^{-}({\bar q})$ and hence the
desired form $b_2^2(q)b_2^2({\bar q})+b_0^2(q)b_0^2({\bar q})$ in
(1.10) is obtained.

\section{Discussion}
\setcounter{equation}{0}
There are many mathematical points of the foregoing computations which
remain to be clarified such as: 1) a direct proof of the
equivalence of the forms (3.13),(3.19) and (4.16) with (2.25), (2.27)
and the forms of appendix B; 2) the obtaining of our results directly
from the functional equations~\cite{Al,Pearce} without recourse to the
study of the completeness rules~\cite{Al2}. However the major feature
of the results of sections~3 and 4 is that they directly relate the
concept of branching function with that of quasi-particle. This
provides insight into the physics of the model which we will discuss
here in detail.

\subsection{Infrared Anomaly}
All the eigenvalue spectra computed here have, up to a possible
constant, the additive quasi-particle form (1.2) of
\beq
E_{ex}-E_{GS}=\sum_{\alpha,rules}e_{\alpha}(P_j),
\eeq
where there are two important features of the rules which govern the
combination of energy levels. The first is the fermi exclusion
property:
\beq
P_i^{\alpha}\neq P_j^{\alpha}\quad {\rm for}\quad i\neq j.
\eeq
This rule in conjunction with the quasi-particle form (5.1) is often
used to say that the quasi-particles are fermions.

If the momenta $P_j^{\alpha}$ were such that
\beq
P_j^{\alpha}={2\pi j\over M}
\eeq
with $j$ an integer (or possibly half integer) these quasi-particle
energies would indeed be identical with those of a genuine free Fermi
gas. However the momentum set out of which the $P_{j}$ are chosen is
not (5.3). Instead the momenta are subject to the restrictions (3.16),
(4.6), (4.13) and (4.14). These all share the feature that there is a
depletion of the number of allowed momenta near the values of $P$
where $e_{\alpha}(P)=0$, which depends
on the number of quasi-particles in the state.
This is an intrinsic many body effect which cannot be modeled by
an effective change in some one or two body property.  Because this
occurs for $e(P)\sim 0$ it seems appropriate to refer to this intrinsic
many body effect as an infrared anomaly. This mechanism is typical of
all conformal field theories that are different from free fermions or
free bosons and is what can lead to central charges being different
from integer or half integer.

This manner of characterizing the
phenomena underlying the behavior of systems with nontrivial branching
functions has, at least on the face of it, nothing to do with
integrability, Virasoro algebra, modular invariance, or any other
symmetry algebra. In
this respect it shares the feature of generality with
Haldane's~\cite{haldane} definition of fractional statistics. However
the definition of~\cite{haldane} differs from that of the infrared
anomaly by relying on the finiteness of the Hilbert space. This can
only be achieved by imposing an ultraviolet as well as an infrared
cutoff on the problem whereas, by  the very name, an infrared anomaly
will exist without an ultraviolet cutoff. The essential feature
of~\cite{haldane}, however, is the abandoning of a second quantized
description of the excitations in the system and this is certainly a
key property of the effects described above.

\subsection{Specific Heat}
One of the powerful results of conformal field theory is the
prediction~\cite{bcn,affleck} that the specific heat $C$, for a system
with periodic boundary condisions, is given in terms of the central
charge and the velocity of sound as $T\sim 0$ as
\beq
C\sim {\pi k_B^2 c\over3v}T.
\eeq
The original derivations are based on conformal invariance. We give
here an argument from the point of view of the infrared anomaly.

By definition the bulk free energy per site is obtained from the partition
function
\beq
Z={\rm Tr}e^{-{H\over k_B T}}
\eeq
as
\beq
f=-k_B T \lim_{M\rightarrow \infty}{1\over M}\ln Z,
\eeq
where
\beq
T>0 \quad {\rm is\quad fixed\quad and}\quad M\rightarrow \infty.
\eeq
Then the specific heat is

\beq
C=-k_BT {\partial^2f\over \partial T^2}.
\eeq

To obtain the leading behavior as $T \rightarrow 0$ of the specific
heat it is sufficient restrict attention to the lowlying (order 1 or
smaller) excitations of $H$ over the ground state. These obey the
quasi particle form (5.1). Hence specific heats are commonly evaluated
with formulae involving single particle levels $e(P)$.  In this
discussion the boundary conditions are involved only in a tacit
fashion.

This argument, however, is not complete, as is apparent in the
observation that any energy level with $e(P)>0$ as $M \rightarrow
\infty$ will contribute only a term exponentially small in $T$ to the
specific heat. Thus the order one excitations do not contribute to the
linear term (5.4). Instead it is the levels which have the property
that $\lim_{M \rightarrow \infty}e(P)=0$ that contribute to the
leading behavior.

The partition function computed in conformal field theory is in the limit
\beq M\rightarrow \infty,\quad T\rightarrow 0\quad{\rm with}\quad
MT\quad {\rm fixed}.\eeq This is not the same as the limit (5.7) which
defines the specific heat. However if no additional length scale
appears in the system it is expected that the behavior of the specific
heat computed using the prescription (5.9) where $q=e^{-{2 \pi v\over
Mk_BT}}$ is fixed will agree when $q \rightarrow 1$ with the $T
\rightarrow 0$ behavior computed using the prescription (5.7).

The $q \rightarrow 1$ behavior of $Z$ can be computed directly from
the expression of the branching functions as infinite series in
$q$. Since these series are a direct consequence of the infrared
anomaly the influence of the many body effects which determine the
momentum selection rules is apparent. These selection rules can only be
seen by imposing an explicit infrared cutoff on the problem. Indeed it
is known~\cite{bcn} that free boundary conditions and periodic
boundary conditions give different amplitudes of the linear in T term
of the specific heat.

The above argument has relied only on a one length scale scaling
argument and has not explicitly used modular invariance. Modular
invariance in this context is a mathematical property of the series
expressions for the branching functions which says that if we
define $\tau$ as $q=e^{2 \pi i \tau}$ then for a set of branching
functions $b_k(\tau)$
\beq
b_k({-1/\tau})=\sum M_{k,l}b_{l}(\tau),
\eeq
where the functions $M_{k,l}$ are independent of $\tau$. Thus the
$q\rightarrow 1$ behavior of the branching functions is given in
terms of the $q \rightarrow 0$ behavior of the branching functions.
This behavior is always $q^{-{c\over 24}+h_k}$ where $h_k$ is the
conformal dimension and c is determined from the finite size
correction to the ground state energy of (3.2). The right hand side of
(5.10) will thus be dominated by the term with the smallest $h_k$. In
particular if the smallest $h_k$ is zero the formula (5.4) results.

This argument has now related the specific heat to the finite size
correction to the ground state energy and here it is apparent by
definition that boundary conditions are important. Nevertheless it is
useful to keep in mind the fact that the specific heat would not be
affected if we changed the finite size corrections provided that the
many body effect of the infrared anomaly was not changed. In that case
the modular invariance would be destroyed but the specific heat would
be unchanged. This is a demonstration that the concept of infrared
anomaly and modular invariance are logically distinct.

\subsection{Oscillations}
One of the striking features of sections~3 and 4 is the fact that the
branching functions $b_2^0$ (with conformal dimension 3/4) and
$b_2^2$ (with conformal dimension 1/12) are obtained with a
macroscopic (order 1) momentum shift from the ground state of $\Delta
P=\pi.$ This results from the feature found in~\cite{Al1} that the
energies $e_{2s}(P)$ $(e_{-2s}(P)$) vanish at $3\pi$ and $\pi$. These
macroscopic momentum shifts are expected to give rise to oscillatory
contributions to the correlation functions of the primary operators of
$b_2^0$ and $b_2^2.$ On the lattice the oscillatory term should be
$(-1)^N$ where $N$ is the separation of the operators. These
microscopic oscillations are, perhaps, unusual for conformal field
theories but are in fact expected if we make the observation that the
anti-ferromagnetic 3-state Potts model lies deep inside the
incommensurate phase of the chiral Potts model~\cite{Al,MR} which is
characterized by oscillatory correlations~\cite{howes,tan}.
\subsection{Lee-Yang Edge}

An interesting property of the branching functions (3.13) and (4.16)
obtains if we consider only the terms where $m_{2s}=m_{-2s}=0$ where
the sums reduce to the two sums
\subequations
\beq
S_0=\sum_{m_{ns}=0}^{\infty}{q^{m_{ns}^2}\over(q)_{m_{ns}}}
\eeq
and
\beq
S_1=\sum_{m_{ns}=0}^{\infty}{q^{m_{ns}(m_{ns}+1)}\over (q)_{m_{ns}}}.
\eeq
\endsubequations
These are the famous sums of Rogers-Ramanujan~\cite{rogers,ram}. They
become modular functions if we multiply by powers of q and consider
\beq
c_{1,3}(\tau )=q^{-{1\over 60}}S_0\quad {\rm and \quad}c_{1,1}
(\tau)=q^{{11\over 60}}S_1,
\eeq
which, in fact, are shown in~\cite{rogers2} to be the two characters of
the nonunitary minimal model~\cite{BPZ} with $p=5$ and $p'=2$
\beq
c_{r,s}(q) = \frac{q^{-1/24}}{Q(q)} \sum_{n=-\infty}^{\infty}
\{q^{(2 n p p' + r p - s p')^2 /4p p'} -
  q^{(2 n p p' + r p + s p')^2 /4 p p'} \}
\eeq
This model has been
identified~\cite{cardy2,isz} with the field theory that describes the
behavior at the Lee-Yang edge of the Ising model~\cite{ly}. Thus there
is a sense in which we may say that the field theory for the Lee-Yang
edge is obtained by adding a perturbation to the 3-state
anti-ferromagnetic Potts chain which makes the $\pm2s$ excitations
massive without affecting the $ns$ excitations.

It is also interesting to note that the $q \rightarrow 1$ behavior of
the sums (5.11) can be calculated directly without recourse to the
modular transformation (5.10) by a method that makes contact with
dilogarithms. As an example we consider explicitly $S_0$ which we
write in the form obtained directly from (3.10)
\beq
S_0=\sum_{m=0}\sum_{n=0}P_d(m,n)q^{n}q^{{m\over 2}(m-1)}
\eeq
(where the subscript ns has been dropped for simplicity). To study the
limit $q \rightarrow 1$ we first use the integral representation
\beq
 \sum_{n=0}^{\infty}P_d(m,n)q^n={1\over 2 \pi i}\oint {dz\over
z^{m+1}} \prod_{l=1}^{\infty}(1+zq^l) \eeq in the sum (5.14). We note
that (5.15) vanishes if $m<0$. Thus we extend the lower limit of the
sum over $m$ from zero to $- \infty$ and interchange the summation and
integration to obtain
\beq
S_0={1\over 2\pi i}\oint {dz\over z}\prod_{l=1}^{\infty}(1+zq^l)
\sum_{m=-\infty}^{\infty}q^{{m\over 2}(m-1)}z^{-m}.
\eeq
The sum over m is expressed in terms of the Jacobi theta
function~\cite{bateman}
\beq
\theta_2(v,q)=\sum_{m=-\infty}^{\infty}q^{(m-{1\over
2})^2}e^{i\pi(2m-1)v}
\eeq
as
\beq
\sum_{m=-\infty}^{\infty}q^{{m\over 2}(m-1)}z^{-m}=
q^{-{1 \over 8}}z^{1\over 2}\theta_{2}(v,q^{1\over 2}),
\eeq
with
\beq
e^{i \pi v}=z^{-{1\over 2}},
\eeq
and hence
\beq
 S_0={1\over 2 \pi i}\oint {dz\over
z}\prod_{l=1}^{\infty}(1+zq^{l})q^{-{1\over 8}}z^{1\over 2}\theta
_2(v,q^{1\over 2}).
\eeq
Then using the product representation
\beq
\theta_2(v,q^{1\over 2})=q^{1\over 8}(z^{1 \over 2}+z^{-{1\over 2}})
\prod_{n=1}^{\infty}(1-q^n)(1+q^n z)(1+q^n z^{-1}),
\eeq
we find
\beq
S_0=\exp\left\{\sum_{n=1}^{\infty}\ln (1-q^n)\right\}
{1\over 2 \pi i}\oint {dz\over
z}(z+1) \exp\left\{\sum_{n=1}^{\infty}
\{2\ln (1+zq^n)+\ln (1+z^{-1}q^n)\}\right\}.
\eeq

We may now study behavior of $S_0$ as $q\rightarrow 1$ by replacing
the sums in (5.22) by integrals. Thus using the definition $q=e^{-{2
\pi v\over M k_B T}}$ and setting
\beq
x={2 \pi v\over M k_B T}
\eeq
we find
\begin{eqnarray}
S_0 &\sim &\exp \left\{{M k_B T \over 2  \pi v}\int_{0}^{\infty}dx
\ln(1-e^{-x})\right\}\nonumber\\
 & &\quad \times{1\over 2\pi i}\oint {dz\over z}
(z+1)\exp\left\{{M k_B T\over 2 \pi v}
\int_{0}^{\infty} dx \{2\ln (1+ze^{-x})+\ln(1+z^{-1}e^{-x})\}\right\} .
\end{eqnarray}

The integral over $z$ may now be evaluated by steepest descents. The
steepest decents point occurs at the values of $z$ that satisfy
\beq
\ln (1+z)^2=\ln (1+z^{-1}),
\eeq
and thus either $z=-1$ or
\beq
1+z=z^{-1},
 \eeq
and hence we find that the steepest descents point is
\beq
z={\sqrt5 -1\over 2}.
\eeq
Thus we have
\beq
S_0\sim \exp\left\{{M k_B T\over 2 \pi v}\int_{0}^{\infty}dx\{\ln
(1-e^{-x}) +2\ln (1+{\sqrt5 -1\over 2}e^{-x})+\ln(1+{\sqrt5 +1\over
2}e^{-x})\}\right\},
\eeq
which, if we define
\beq
t=e^{-x},
\eeq
and recall the definition~\cite{lewin} of the dilogarithm
\beq
{\rm Li}_2(z)=-\int_{0}^{z}dt {\ln(1-t)\over t},
\eeq
may be rewritten as
\beq
S_0\sim \exp\left\{-{M k_B T\over 2\pi v}\{{\rm Li}_2(1)
+2{\rm Li}_2({1-\sqrt5\over 2})+{\rm Li}_2(-{\sqrt5 +1\over 2})\}\right\}.
\eeq
Then if we note the special values of the dilogarithm
\subequations
\beq
{\rm Li}_2(1)={\pi^2\over 6},
\eeq
\beq
{\rm Li}_2({1-\sqrt5\over 2})=-{\pi^2\over 15}+{1\over 2}\ln^2({\sqrt5
-1\over 2}),
\eeq
\beq
{\rm Li}_2(-{1+ \sqrt5\over 2})=-{\pi^2\over 10}-\ln^2({\sqrt5 +1\over 2}),
\eeq
\endsubequations
we obtain the result
\beq
S_0\sim \exp{M k_B T \pi \over 30 v}.
\eeq
The identical behavior is obtained for $S_1$.  Thus with $Z=
S_0(q)S_0(\bar q) + S_1(q)S_1(\bar q)$, we obtain from (5.4)-(5.8) an
(effective) central charge of 2/5 which agrees with~\cite{isz}.

\section*{Acknowledgements}

We are pleased to acknowledge many useful discussions with Dr. G.
Albertini, Prof. R. J. Baxter, Prof. V. V. Bazhanov, Prof. J. L.
Cardy, Dr. S. Dasmahapatra, and Prof. C. Itzykson.
One of us (BMM) is pleased to thank Prof. P. A. Pearce for many
explanations of conformal field theory and
for hospitality
extended at the University of Melbourne where this work was started
and to thank Prof. R. J.  Baxter for hospitality extended at the Isaac
Newton Institute.
We are indebted to Dr. E. Melzer for the argument of appendix A.
 This work was partially supported by the National
Science Foundation under grant DMR-9106648.

\begin{appendix}

\section*{Appendix A. Gaussian construction of the branching functions}
\setcounter{section}{1}
\setcounter{equation}{0}N
The form of the branching function (2.27) may be simply obtained if
we note that the $Z_{4}$ parafermions  with the diagonal ($A_5$) modular
invariant partition function is known to be the $r=\sqrt{3/2}$ point
on the orbifold line of $c=1$ conformal field theories (see
e.g.~\cite{ginsparg}).  However, we are interested in the nondiagonal
($D_4$) partition function which contains the current operator of
dimension (1,0) in the spectrum.  This model must therefore lie on the
$c=1$ Gaussian line at the compactification radius $r=\sqrt{3/2}$.
Indeed, consider the general Gaussian model partition function
\beq
Z(r)={1\over \eta(q)\eta({\bar q})}\sum_{m,n=-\infty}^{\infty}
q^{\Delta_{m,n}}{\bar q}^{\bar \Delta_{m,n}},
\eeq
where
\beq
\Delta_{m,n}(r)={1\over2}({m\over 2r}+nr)^2\quad {\rm and}\quad
\bar \Delta_{m,n}(r)={1\over 2}({m\over 2r}-nr)^2
\eeq
and set $r={\sqrt{3/2}}$ to obtain
\beq
Z(\sqrt{\frac{3}{2}})={1\over \eta(q)\eta({\bar q})}\sum_{m,n=-\infty}^\infty
q^{3({m+3n\over 6})^2} {\bar q}^{3({m-3n\over 6})^2}.
\eeq
Rewriting the sum as
\beq
Z(\sqrt{\frac{3}{2}})={1\over\eta(q)\eta({\bar q})}\sum_{b=0}^5
\sum_{\stackrel{m,n=-\infty}{m+3n\equiv b\,{\rm mod\ }6}}^\infty
q^{3({m+3n\over 6})^2}{\bar q}^{3({m-3n\over 6})^2}
\eeq
we see that
\beq
Z(\sqrt{\frac{3}{2}})={1\over \eta(q)\eta({\bar q})}
\sum_{b=0}^5f_{3,b}(q)f_{3,b}({\bar q})
\eeq
with $f_{m,n}$ defined by (2.29). Then noting the symmetry
\beq
f_{3,1}=f_{3,5}\quad {\rm and}\quad f_{3,2}=f_{3,4}
\eeq
we obtain precisely the $Z_{pf4}$ of (1.10) with the expressions of
(2.27) for $b_0^0+b_4^0$, $b_2^0$,
$b_0^2$, and $b_2^2$.

\section*{Appendix B: Branching functions of $\bf
(A_3^{(1)})_1\times(A_3^{(1)})_1/(A_3^{(1)})_2$}
\setcounter{section}{2}
\setcounter{equation}{0}
The branching functions (2.27) can also be expressed as a three
dimensional sum in the following way~\cite{jimbo,cnr,fy2}.  Let
$\alpha_i$, $i=1,2,3$ be the simple roots of $A_3$.  Then
\beq
(\alpha_i,\alpha_j) = c_{i,j},
\eeq
where $c$ is the Cartan matrix of $A_3$
\beq c = \left(\matrix{2&-1&0\cr
                       -1&2&-1\cr
                       0&-1&2\cr}\right). \eeq
Let $\lambda_i$ be the fundamental weights of $A_3$, so that
$\lambda_j =\sum_k c_{j,k}^{-1} \alpha_k$.
The dominant weights
${\bf a}^{(k)}$ of level $k$ are defined to be:
\beq
{\bf a}^{(k)} = \sum_{i=1}^3 a_i \lambda_i, \ \ \
                           a_i\in {\bf Z},\ \ a_i \geq 0,\ \
\sum_{i=1}^3 a_i  \leq k.
\eeq
Let ${\bf r = a}^{(1)} + \rho$, ${\bf s= a}^{(2)} + \rho$,
 where $\rho = \sum_i\lambda_i$.
Then the following branching functions  are identical with (2.27):
\beq
b_{\bf r,s} = \frac{1}{\eta^3} \sum_{\bf k\in Q} \sum_{w\in W}
\det(w) q^{\frac{| 30 {\bf k} - 6{\bf r }+ 5 w({\bf s})|^2}{60}}
\eeq
where $W$ is the Weyl group of  $A_3$, generated by the three simple
reflections:
\beq
\sigma_i (\beta) = \beta - (\alpha_i,\beta) \alpha_i, \ \ \ i=1,2,3
\eeq
This group has 24 elements made up of powers of the simple
reflections, and $\det(w)=\pm1$, depending on whether the minimum number of
simple reflections making up the element $w$ is even or odd.
$\bf Q$ is the root lattice of $A_3$, i.e. ${\bf k}
=\sum_{i}m_i \alpha_i$, and the
sum $\sum_{\bf k \in Q}$ is thus a sum over $m_i, i=1,2,3$, from
$-\infty$ to $\infty$.

Equation (B.4) gives 7 unique branching functions, but only 5 appear in our
model: these can be obtained by setting $\bf r = \rho$ and choosing
the following ${\bf a}^{(2)} = {\bf s} - \rho$:
\beq
{\bf a}^{(2)} = \cases{ 0, & $b_{\bf r, s} = b_0^0,\ h_{\bf r, s}=0$ \cr
                  2\lambda_2, & $b_{\bf r, s} = b_4^0,\ h_{\bf r, s}=1$\cr
                  2\lambda_1, & $b_{\bf r, s} = b_2^0,\ h_{\bf r, s}=3/4$\cr
                  \lambda_2,  & $b_{\bf r, s} = b_2^2,\ h_{\bf r, s}=1/12$\cr
                  \lambda_1+\lambda_3, & $b_{\bf r, s} = b_0^2,
                                             \ h_{\bf r, s}=1/3$.\cr}
\eeq
Here, the conformal dimension $h_{\bf r, s}$ is defined to be:
\beq
h_{\bf r, s} = -\frac{1}{12} + \frac{|-6{\bf r} + 5 {\bf s}|^2}{60}.
\eeq

The three dimensional sum (B.4) is to be compared with the expressions
(3.13) and (4.16) of the text. We note that the sum (B.4) has a power
$q^{\sum 30 m_{j}^2}$ and thus the powers of $q$ grow much more rapidly
as a function of $m_{j}$ than do (3.13) and (4.16). Note also that
when the sum over $W$ is performed (B.4) contains 24 triple sums whereas
(3.13) has only one and (4.16) has two.

\end{appendix}

\begin{table}
\begin{center}
\caption{The terms through order $q^8$ in the construction of
$b_0^0$ from the rules of section~3. The minimum momenta
$P_{min}^{ns}={\pi\over M}(m_{ns}+m_{2s}+m_{-2s}+1)$ and $P_{ns}^{\pm 2s}=
{\pi \over M}(m_{ns}+{m_{2s}+m_{-2s}\over 2}+1)$ are obtained
from (3.15). The terms in $q^{1/24}b_0^0$ are obtained from (2.30a).
Here $m_{2s}+m_{-2s}$ and $m_{ns}+m_{-2s}+{m_{2s}+m_{-2s}\over 2}$ are even.
The macroscopic momentum shift is $\Delta P = 0$.}

\begin{tabular}{|l|l|l|l|l|l|l|l|l|} \hline
order & $m_{ns}^\ell$ & $m_{2s}^\ell$ & $m_{-2s}^\ell$ &
$P_{min}^{ns}$&$P_{min}^{\pm2s}$&
$\{P^{ns},P^{2s},P^{-2s}\}$ (units of $\frac{\pi}{M}$) &states &
$q^{1/24}b_0^0$ \\ \hline
$q^0$ & 0 & 0 & 0 &$-$& $-$& $\{0,0,0\}$  & 1 & 1 \\ \hline
$q^2$ & 0 & 1 & 1 & $-$ & $2\pi/M$ & $\{0,2,2\}$ & 1 & 1 \\ \hline
$q^3$ & 0 & 1 & 1 & $-$ & $2\pi/M$ & $\{0,4,2\},\{0,2,4\}$&2&2\\ \hline
$q^4$ & 0 & 1 & 1 & $-$ & $2\pi/M$ & $\{0,6,2\},\{0,4,4\},\{0,2,6\}$&3&\\
      & 2 & 0 & 0 & $3\pi/M$ & $-$ & $\{3+5,0,0\}$ & 1 &4 \\ \hline
$q^5$ & 0 & 1 & 1 & $-$ & $2\pi/M$ & $\{0,8,2\},\{0,6,4\},\{0,4,6\},$& &\\
      &   &   &   &     &          & $\{0,2,8\}$ & 4 & \\
      & 2 & 0 & 0 & $3\pi/M$ & $-$ & $\{3+7,0,0\}$ & 1 &5 \\ \hline
$q^6$ & 0 & 1 & 1 & $-$ & $2\pi/M$ & $\{0,10,2\},\{0,8,4\},\{0,6,6\},$& & \\
      &   &   &   &   &                  & $\{0,4,8\},\{0,2,10\}$ & 5 &  \\
      & 2 & 0 & 0 & $3\pi/M$ & $-$ & $\{3+9,0,0\},\{5+7,0,0\} $ & 2& \\
      & 1 & 2 & 0 & $4\pi/M$ & $3\pi/M$ & $\{4,3+5,0\}$& 1 & \\
      & 1 & 0 & 2 & $4\pi/M$ & $3\pi/M$ & $\{4, 0, 3+5\}$ & 1&9 \\ \hline
$q^7$ & 0 & 1 & 1 & $-$ & $2\pi/M$ & $\{0,12,2\},\{0,10,4\},\{0,8,6\},$& &\\
      &   &   &   &   &                  & $\{0,6,8\},\{0,4,10\},\{0,2,12\} $ &
6 & \\
      & 2 & 0 & 0 & $3\pi/M$ & $-$ & $\{3+11,0,0\},\{5+9,0,0\} $ & 2 & \\
      & 1 & 2 & 0 & $4\pi/M$&$3\pi/M$& $\{6,3+5,0\},\{4,3+7\}$ & 2 &\\
      & 1 & 0 & 2 & $4\pi/M$&$3\pi/M$& $\{6, 0, 3+5\},\{4,0,3+7\}$& 2 &12 \\
\hline
$q^8$ & 0 & 1 & 1 & $-$ & $2\pi/M$ & $\{0,14,2\},\{0,12,4\},\{0,10,6\}$& & \\
      &   &   &   &     &          & $\{0,8,8\}, \{0,6,10\},\{0,4,12\}$& & \\
      &   &   &   &     &          & $\{0,2,14\}$&7 & \\
      & 2 & 0 & 0 & $3\pi/M$ & $-$ & $\{3+13,0,0\},\{5+11,0,0\},$&  & \\
      &   &   &   &          &     & $\{7+9,0,0\}$ &3&\\
      & 1 & 2 & 0 & $4\pi/M$ & $3\pi/M $ & $\{8,3+5,0\},\{6,3+7,0\},$ & &\\
      &   &   &   &          &           & $\{4,3+9,0\},\{4,5+7,0\} $&4 & \\
      & 1 & 0 & 2 & $4\pi/M$ & $3\pi/M $ & $\{8,0,3+5\},\{6,0,3+7\}$ & & \\
      &   &   &   &          &           & $\{4,0,3+9\},\{4,,0,5+7\}$ &4 & \\
      & 0 & 2 & 2 & $-$ & $3\pi/M$ & $\{0,3+5,3+5\}$ & 1 & 19 \\ \hline
\end{tabular}
\end{center}
\label{table1}
\end{table}

\setcounter{table}{1}
\begin{table}
\begin{center}
\caption{The terms through order $q^{31/4}$ in the construction of
$b_2^0$ from the rules of section 3. The minimum momenta
$P_{min}^{ns}={\pi\over M}(m_{ns}+m_{2s}+m_{-2s}+1)$ and $P_{min
}^{\pm2s}={\pi \over M}(m_{ns}+{m_{2s}+m_{-2s}\over 2}+1)$ are obtained
from (3.15). The terms in $q^{1/24}b_2^0$ are obtained from (2.30b).
Here $m_{2s}+m_{-2s}$ is odd and $m_{2s}<m_{-2s}$.
The macroscopic momentum shift is $\Delta P = \pi$.}

\begin{tabular}{|l|l|l|l|l|l|l|l|l|} \hline
order & $m_{ns}^\ell$ & $m_{2s}^\ell$ & $m_{-2s}^\ell$ &
$P_{min}^{ns}$&$P_{min}^{\pm2s}$&
$\{P^{ns},P^{2s},P^{-2s}\}$ (units of $\frac{\pi}{M}$) &states &
$q^{1/24}b_2^0$ \\ \hline
$q^{3/4}$ & 0 & 0 & 1 & $-$ & $3\pi/2M$ & $\{0,0,3/2\}$ & 1 & 1\\ \hline
$q^{7/4}$ & 0 & 0 & 1 & $-$ & $3\pi/2M$ & $\{0,0,7/2\}$ & 1 & 1\\ \hline
$q^{11/4}$ & 0 & 0 & 1 & $-$ & $3\pi/2M$ & $\{0,0,11/2\}$ & 1 & \\
         & 1 & 0& 1 & $3\pi/M$ & $5\pi/2M$ & $\{3,0,5/2\}$ & 1 & 2 \\ \hline
$q^{15/4}$ & 0 & 0 & 1 & $-$ & $3\pi/2M$ & $\{0,0,15/2\}$ & 1 & \\
 & 1 & 0 &1 & $3\pi/M$ & $5\pi/2M$ & $\{3,0,9/2\},\{5,0,5/2\}$ & 2 & 3 \\
\hline
$q^{19/4}$ & 0 & 0 & 1 & $-$ & $3\pi/2M$ & $\{0,0,19/2\}$ & 1 & \\
 & 1 & 0&1&$3\pi/M$&$5\pi/2M$&$\{3,0,13/2\},\{5,0,9/2\},\{7,0,5/2\}$ & 3 & \\
 & 0 & 1 & 2 &$-$ & $5\pi/2M$ & $\{0,5/2,5/2+9/2\}$ & 1 & 5 \\ \hline
$q^{23/4}$ & 0 & 0 & 1 & $-$ & $3\pi/2M$ & $\{0,0,23/2\}$ & 1 & \\
 & 1 & 0&1&$3\pi/M$&$5\pi/2M$&$\{3,0,17/2\},\{5,0,13/2\}$ &  & \\
 &   &  & &        &         &$\{7,0,9/2\},\{9,0,5/2\}$ & 4 & \\
 & 0 & 1 & 2 &$-$ & $5\pi/2M$ & $\{0,9/2,5/2+9/2\},\{0,5/2,5/2+13/2\}$ & 2 & 7
\\ \hline
$q^{27/4}$ & 0 & 0 & 1 & $-$ & $3\pi/2M$ & $\{0,0,27/2\}$ & 1 & \\
 &1&0&1&$3\pi/M$&$5\pi/2M$&$\{3,0,21/2\},\{5,0,17/2\},\{7,0,13/2\}$& & \\
 & & & &        &         &$\{9,0,9/2\},\{11,0,5/2\}$&5 & \\
 &0&1&2&$-$&$5\pi/2M$&$\{0,13/2,5/2+9/2\},\{0,9/2,5/2+13/2\}$& & \\
 & & & &   &         &$\{0,5/2,5/2+17/2\},\{0,5/2,9/2+13/2\}$&4 & \\
 &2&0&1&$4\pi/M$&$7\pi/2M$&$\{4+6,0,7/2\}$&1& \\
 &0&0&3&$-$&$5\pi/2M$&$\{0,0,5/2+9/2+13/2\}$&1&12\\ \hline
$q^{31/4}$ & 0 & 0 & 1 & $-$ & $3\pi/2M$ & $\{0,0,31/2\}$ & 1 & \\
 &1&0&1&$3\pi/M$&$5\pi/2M$&$\{3,0,25/2\},\{5,0,21/2\},\{7,0,17/2\}$& &\\
 & & & &        &         &$\{9,0,13/2\},\{11,0,9/2\},\{13,0,5/2\}$&6 &\\
 &0&1&2&$-$&$5\pi/2M$&$\{0,5/2,5/2+21/2\},\{0,5/2,9/2+17/2\}$& & \\
 & & & &   &         &$\{0,9/2,5/2+17/2\},\{0,9/2,9/2+13/2\}$& & \\
 & & & &   &         &$\{0,13/2,5/2+13/2\},\{0,17/2,5/2+9/2\}$&6 & \\
 &2&0&1&$4\pi/M$&$7\pi/2M$&$\{4+8,0,7/2\},\{4+6,0,11/2\}$&2&\\
 &0&0&3&$-$&$5\pi/2M$&$\{0,0,5/2+9/2+17/2\}$& 1 & 16 \\ \hline
\end{tabular}
\end{center}
\end{table}

\setcounter{table}{2}
\begin{table}
\begin{center}
\caption{The terms through order $q^8$ in the construction of
$b_4^0$ from the rules of section 3. The minimum momenta
$P_{min}^{ns}={\pi\over M}(m_{ns}+m_{2s}+m_{-2s}+1)$ and $P_{min
}^{\pm}={\pi \over M}(m_{ns}+{m_{2s}+m_{-2s}\over 2}+1)$ are obtained
from (3.15). The terms in $q^{1/24}b_4^0$ are obtained from (2.30c).
Here $m_{2s}+m_{-2s}$ is even and
$m_{ns}+m_{-2s}+{m_{2s}+m_{-2s}\over 2}$ is odd.
The macroscopic momentum shift is $\Delta P = 0$.}

\begin{tabular}{|l|l|l|l|l|l|l|l|l|} \hline
order & $m_{ns}^\ell$ & $m_{2s}^\ell$ & $m_{-2s}^\ell$ &
$P_{min}^{ns}$&$P_{min}^{\pm2s}$&
$\{P^{ns},P^{2s},P^{-2s}\}$ (units of $\frac{\pi}{M}$) &states &
$q^{1/24}b_4^0$ \\ \hline
$q^1$ & 1 & 0 & 0 & $2\pi/M$ & $-$ & $\{2,0,0\}$ & 1 & 1 \\ \hline
$q^2$ & 1 & 0 & 0 & $2\pi/M$ & $-$ & $\{4,0,0\}$ & 1 & 1 \\ \hline
$q^3$ & 1 & 0 & 0 & $2\pi/M$ & $-$ & $\{6,0,0\}$ & 1 &   \\
      & 0 & 2 & 0 & $-$ & $2\pi/M$ & $\{0,2+4,0\}$&1 &   \\
      & 0 & 0 & 2 & $-$ & $2\pi/M$ & $\{0,0,2+4\}$&1 & 3 \\ \hline
$q^4$ & 1 & 0 & 0 & $2\pi/M$ & $-$ & $\{8,0,0\}$ & 1 &   \\
      & 0 & 2 & 0 & $-$ & $2\pi/M$ & $\{0,2+6,0\}$&1 &   \\
      & 0 & 0 & 2 & $-$ & $2\pi/M$ & $\{0,0,2+6\}$&1 & 3 \\ \hline
$q^5$ & 1 & 0 & 0 & $2\pi/M$ & $-$ & $\{10,0,0\}$ & 1 &   \\
      & 0 & 2 & 0 & $-$ & $2\pi/M$ & $\{0,2+8,0\},\{0,4+6,0\}$&2 & \\
      & 0 & 0 & 2 & $-$ & $2\pi/M$ & $\{0,0,2+8\},\{0,0,4+6\}$&2 & \\
      & 1 & 1 & 1 & $4\pi/M$ & $3\pi/M$ & $\{4,3,3\}$&1 & 6 \\ \hline
$q^6$ & 1 & 0 & 0 & $2\pi/M$ & $-$ & $\{12,0,0\}$ & 1 &   \\
      & 0 & 2 & 0 & $-$ & $2\pi/M$ & $\{0,2+10,0\},\{0,4+8,0\}$&2 & \\
      & 0 & 0 & 2 & $-$ & $2\pi/M$ & $\{0,0,2+10\},\{0,0,4+8\}$&2 & \\
      & 1 & 1 & 1 & $4\pi/M$ & $3\pi/M$ & $\{4,5,3\},\{4,3,5\},\{6,3,3\}$&3 & 8
\\ \hline
$q^7$ & 1 & 0 & 0 & $2\pi/M$ & $-$ & $\{14,0,0\}$ & 1 &   \\
  & 0 & 2 & 0 & $-$ & $2\pi/M$ & $\{0,2+12,0\},\{0,4+10,0\},$& & \\
  &   &   &   &     &          & $\{0,6+8,0\}$&3& \\
  & 0 & 0 & 2 & $-$ & $2\pi/M$ & $\{0,0,2+12\},\{0,0,4+10\},$& & \\
  &   &   &   &     &          & $\{0,0,6+8\}$ & 3 & \\
  & 1 & 1 & 1 & $4\pi/M$ & $3\pi/M$ & $\{4,7,3\},\{4,5,5\},\{4,3,7\}$& &\\
  &   &   &   &          &          & $\{6,3,5\},\{6,5,3\},\{8,3,3\}$& 6 & 13\\
 \hline
$q^8$ & 1 & 0 & 0 & $2\pi/M$ & $-$ & $\{16,0,0\}$ & 1 &   \\
  & 0 & 2 & 0 & $-$ & $2\pi/M$ & $\{0,2+14,0\},\{0,4+12,0\},$& & \\
  &   &   &   &     &          & $\{0,6+10,0\}$ & 3 & \\
  & 0 & 0 & 2 & $-$ & $2\pi/M$ & $\{0,0,2+14\},\{0,0,4+12\},$& & \\
  &   &   &   &     &          & $\{0,0,6+10\}$ & 3 & \\
  & 1 & 1 & 1 & $4\pi/M$&$3\pi/M$& $\{4,9,3\},\{4,7,5\},\{4,5,7\}$& &\\
  &   &   &   &          &       & $\{4,3,9\},\{6,7,3\},\{6,5,5\}$&  &\\
  &   &   &   &          &       & $\{6,3,7\},\{8,5,3\},\{8,3,5\}$&  &\\
  &   &   &   &          &       & $\{10,3,3\}$&10  &17\\
 \hline
\end{tabular}
\end{center}
\end{table}

\setcounter{table}{3}
\begin{table}
\begin{center}
\caption{The terms through order $q^8$ in the sector
$m_{-+}-m_{++}=-1$ constructed from the rules of section 4.1.
The minimum momenta $P_{min}^{ns}={\pi\over
M}(m_{ns}+m_{2s}+m_{-2s}+1)$ and $P_{min}^{\pm2s}={\pi\over
M}(m_{ns}+{m_{2s}+m_{-2s}\over 2}+2)$ are obtained from (4.6). Here
$m_{2s}+m_{-2s}$ is even and the macroscopic momentum shift is $\Delta P=\pi$.}

\begin{tabular}{|l|l|l|l|l|l|l|l|l|} \hline
order & $m_{ns}^\ell$ & $m_{2s}^\ell$ & $m_{-2s}^\ell$ &
$P_{min}^{ns}$&$P_{min}^{\pm2s}$&
$\{P^{ns},P^{2s},P^{-2s}\}$ (units of $\frac{\pi}{M}$) &states & total \\
\hline
$q^0$ & 0 & 0 & 0 & $-$ & $-$ & $\{0,0,0\}$ & 1 & 1 \\ \hline
$q^1$ & 1 & 0 & 0 & $2\pi/M$ & $-$ & $\{2,0,0\}$ & 1 & 1 \\ \hline
$q^2$ & 1 & 0 & 0 & $2\pi/M$ & $-$ & $\{4,0,0\}$ & 1 & 1 \\ \hline
$q^3$ & 1 & 0 & 0 & $2\pi/M$ & $-$ & $\{6,0,0\}$ & 1 &  \\
      & 0 & 1 & 1 & $-$ & $3\pi/M$ & $\{0,3,3\}$ & 1 & 2 \\ \hline
$q^4$ & 1 & 0 & 0 & $2\pi/M$ & $-$ & $\{8,0,0\}$ & 1 &  \\
      & 0 & 1 & 1 & $-$ & $3\pi/M$ & $\{0,3,5\},\{0,5,3\}$ & 2 &  \\
      & 2 & 0 & 0 & $3\pi/M$ & $-$ & $\{3+5,0,0\}$ & 1 & \\
      & 0 & 2 & 0 & $-$ & $3\pi/M$ & $\{0,3+5,0\}$ & 1 & \\
      & 0 & 0 & 2 & $-$ & $3\pi/M$ & $\{0,0,3+5\}$ & 1 & 6\\ \hline
$q^5$ & 1 & 0 & 0 & $2\pi/M$ & $-$ & $\{10,0,0\}$ & 1 &  \\
      & 0 & 1 & 1 & $-$ & $3\pi/M$ & $\{0,3,7\},\{0,5,5\},\{0,7,3\}$ & 3 &  \\
      & 2 & 0 & 0 & $3\pi/M$ & $-$ & $\{3+7,0,0\}$ & 1 & \\
      & 0 & 2 & 0 & $-$ & $3\pi/M$ & $\{0,3+7,0\}$ & 1 & \\
      & 0 & 0 & 2 & $-$ & $3\pi/M$ & $\{0,0,3+7\}$ & 1 & 7\\ \hline
$q^6$ & 1 & 0 & 0 & $2\pi/M$ & $-$ & $\{12,0,0\}$ & 1 &  \\
      & 0 & 1 & 1 & $-$ & $3\pi/M$ & $\{0,3,9\},\{0,5,7\},$ &  &  \\
      &   &   &   &     &          & $\{0,7,5\},\{0,9,3\} $ & 4& \\
      & 2 & 0 & 0 & $3\pi/M$ & $-$ & $\{3+9,0,0\},\{5+7,0,0\}$ & 2 & \\
      & 0 & 2 & 0 & $-$ & $3\pi/M$ & $\{0,3+9,0\},\{0,5+7,0\}$ & 2 & \\
      & 0 & 0 & 2 & $-$ & $3\pi/M$ & $\{0,0,3+7\},\{0,0,5+7\}$ & 2 & \\
      & 1 & 1 & 1 & $4\pi/M$ & $4\pi/M$ & $\{4,4,4\}$ & 1 & 12 \\ \hline
$q^7$ & 1 & 0 & 0 & $2\pi/M$ & $-$ & $\{14,0,0\}$ & 1 &  \\
      & 0 & 1 & 1 & $-$ & $3\pi/M$ & $\{0,3,11\},\{0,5,9\},\{0,7,7\},$ &  &  \\
      &   &   &   &     &          & $\{0,9,5\},\{0,11,3\} $ & 5& \\
      & 2 & 0 & 0 & $3\pi/M$ & $-$ & $\{3+11,0,0\},\{5+9,0,0\}$ & 2 & \\
      & 0 & 2 & 0 & $-$ & $3\pi/M$ & $\{0,3+11,0\},\{0,5+9,0\}$ & 2 & \\
      & 0 & 0 & 2 & $-$ & $3\pi/M$ & $\{0,0,3+11\},\{0,0,5+9\}$ & 2 & \\
      & 1 & 1 & 1 & $4\pi/M$ & $4\pi/M$ & $\{4,4,6\},\{4,6,4\},\{6,4,4\}$ & 3 &
\\
      & 1 & 2 & 0 & $4\pi/M$ & $4\pi/M$ & $\{4,4+6,0\}$ & 1 & \\
      & 1 & 0 & 2 & $4\pi/M$ & $4\pi/M$ & $\{4,0,4+6\}$ & 1 & 17\\ \hline
$q^8$ & 1 & 0 & 0 & $2\pi/M$ & $-$ & $\{16,0,0\}$ & 1 &  \\
      & 0 & 1 & 1 & $-$ & $3\pi/M$ & $\{0,3,13\},\{0,5,11\},\{0,7,9\},$ &  &
\\
      &   &   &   &     &          & $\{0,9,7\},\{0,11,5\},\{0,13,3\} $ & 6& \\
      & 2 & 0 & 0 & $3\pi/M$ & $-$ & $\{3+13,0,0\},\{5+11,0,0\},\{7+9,0,0\}$ &
3 & \\
      & 0 & 2 & 0 & $-$ & $3\pi/M$ & $\{0,3+13,0\},\{0,5+11,0\},\{0,7+9,0\}$ &
3 & \\
      & 0 & 0 & 2 & $-$ & $3\pi/M$ & $\{0,0,3+13\},\{0,0,5+11\},\{0,0,7+9\}$ &
3 & \\
      & 1 & 1 & 1 & $4\pi/M$ & $4\pi/M$ & $\{4,4,8\},\{4,8,4\},\{8,4,4\},$ &  &
\\
      &   &   &   &          &          & $\{4,6,6\},\{6,4,6\},\{6,6,4\}$  & 6
& \\
      & 1 & 2 & 0 & $4\pi/M$ & $4\pi/M$ & $\{4,4+8,0\},\{6,4+6,0\}$ & 2 & \\
      & 1 & 0 & 2 & $4\pi/M$ & $4\pi/M$ & $\{4,0,4+8\},\{6,0,4+6\}$ & 2 & 26\\
\hline
\end{tabular}
\end{center}
\end{table}

\setcounter{table}{4}
\begin{table}
\begin{center}
\caption{The terms through order $q^8$ in the sector
$m_{-+}-m_{++}=1$ constructed from the rules of section 4.2.
The minimum momenta $P_{min}^{ns}={\pi\over
M}(m_{ns}+m_{2s}+m_{-2s}+3)$ and $P_{min}^{\pm2s}={\pi\over
M}(m_{ns}+{m_{2s}+m_{-2s}\over 2}+3)$ are obtained from (4.10). Here
$m_{2s}+m_{-2s}$ is even and the macroscopic momentum shift is $\Delta P=0$.}

\begin{tabular}{|l|l|l|l|l|l|l|l|l|} \hline
order & $m_{ns}^\ell$ & $m_{2s}^\ell$ & $m_{-2s}^\ell$ &
$P_{min}^{ns}$&$P_{min}^{\pm2s}$&
$\{P^{ns},P^{2s},P^{-2s}\}$ (units of $\frac{\pi}{M}$) &states & total \\
\hline
$q^0$ & 0 & 0 & 0 & $-$ & $-$ & $\{0,0,0\}$ & 1 & 1\\ \hline
$q^2$ & 1 & 0 & 0 & $4\pi/M$ & $-$ & $\{4,0,0\}$ & 1 & 1 \\ \hline
$q^3$ & 1 & 0 & 0 & $4\pi/M$ & $-$ & $\{6,0,0\}$ & 1 & 1 \\ \hline
$q^4$ & 1 & 0 & 0 & $4\pi/M$ & $-$ & $\{8,0,0\}$ & 1 &   \\
      & 0 & 1 & 1 & $-$ & $4\pi/M$ & $\{0,4,4\}$ & 1 & 2 \\ \hline
$q^5$ & 1 & 0 & 0 & $4\pi/M$ & $-$ & $\{10,0,0\}$ & 1 &   \\
      & 0 & 1 & 1 & $-$ & $4\pi/M$ & $\{0,4,6\},\{0,6,4\}$ & 2 &  \\
      & 0 & 2 & 0 & $-$ & $4\pi/M$ & $\{0,4+6,0\}$ & 1 &  \\
      & 0 & 0 & 2 & $-$ & $4\pi/M$ & $\{0,0,4+6\}$ & 1 & 5 \\ \hline
$q^6$ & 1 & 0 & 0 & $4\pi/M$ & $-$ & $\{12,0,0\}$ & 1 &   \\
      & 0 & 1 & 1 & $-$ & $4\pi/M$ & $\{0,4,8\},\{0,6,6\},\{0,8,4\}$ & 3 &  \\
      & 0 & 2 & 0 & $-$ & $4\pi/M$ & $\{0,4+8,0\}$ & 1 &  \\
      & 0 & 0 & 2 & $-$ & $4\pi/M$ & $\{0,0,4+8\}$ & 1 &   \\
      & 2 & 0 & 0 & $5\pi/M$ & $-$ & $\{5+7,0,0\}$ & 1 & 7 \\ \hline
$q^7$ & 1 & 0 & 0 & $4\pi/M$ & $-$ & $\{14,0,0\}$ & 1 &   \\
      & 0 & 1 & 1 & $-$ & $4\pi/M$ & $\{0,4,10\},\{0,6,8\},$ &  &  \\
      &   &   &   &     &          & $\{0,8,6\},\{0,10,4\}$ & 4 & \\
      & 0 & 2 & 0 & $-$ & $4\pi/M$ & $\{0,4+10,0\},\{0,6+8,0\},$ & 2 &  \\
      & 0 & 0 & 2 & $-$ & $4\pi/M$ & $\{0,0,4+10\},\{0,0,6+8\}$ & 2 &   \\
      & 2 & 0 & 0 & $5\pi/M$ & $-$ & $\{5+9,0,0\}$ & 1 & 10 \\ \hline
$q^8$ & 1 & 0 & 0 & $4\pi/M$ & $-$ & $\{16,0,0\}$ & 1 &   \\
      & 0 & 1 & 1 & $-$ & $4\pi/M$ & $\{0,4,12\},\{0,6,10\},\{0,8,8\},$ &  &
\\
      &   &   &   &     &          & $\{0,10,6\},\{0,12,4\}$ & 5 & \\
      & 0 & 2 & 0 & $-$ & $4\pi/M$ & $\{0,4+12,0\},\{0,6+10,0\},$ & 2 &  \\
      & 0 & 0 & 2 & $-$ & $4\pi/M$ & $\{0,0,4+12\},\{0,0,6+10\}$ & 2 &   \\
      & 2 & 0 & 0 & $5\pi/M$ & $-$ & $\{5+11,0,0\},\{7+9,0,0\}$ & 2 & \\
      & 1 & 1 & 1 & $6\pi/M$ & $5\pi/M$ & $\{6,5,5\}$ & 1 & 13 \\ \hline
\end{tabular}
\end{center}
\end{table}

\setcounter{table}{5}
\begin{table}
\begin{center}
\caption{The terms through order $q^8$ in the sector
$m_{-+}=m_{++}$ with the macroscopic momentum $\Delta P=0$ constructed
from the rules of sec, 4.3. The minimum momenta
$P_{min}^{ns}={\pi\over M}(m_{ns}+m_{2s}+m_{-2s}+1)$ and $P_{min
}^{\pm2s}={\pi\over M}(m_{ns}+{m_{2s}+m_{-2s}+1\over 2}+1)$ are
obtained from (4.13). Here $m_{2s}+m_{-2s}$ is odd and only
$m_{2s}>m_{-2s}$ are explicitly shown.}
\begin{tabular}{|l|l|l|l|l|l|l|l|l|l|} \hline
order & $m_{ns}^\ell$ & $m_{2s}^\ell$ & $m_{-2s}^\ell$ &
$P_{min}^{ns}$&$P_{min}^{\pm2s}$&
$\{P^{ns},P^{2s},P^{-2s}\}$ (units of $\frac{\pi}{M}$)&shift &states & total \\
\hline
$q^1$ & 0 & 1 & 0 & $-$ & $2\pi/M$ & $\{0,2,0\}$ & 0 & 1 &2 \\ \hline
$q^2$ & 0 & 1 & 0 & $-$ & $2\pi/M$ & $\{0,4,0\}$ & 0 & 1 & 2\\ \hline
$q^3$ & 0 & 1 & 0 & $-$ & $2\pi/M$ & $\{0,6,0\}$ & 0 & 1 & \\
      & 1 & 1 & 0 & $3\pi/M$ & $3\pi/M$ & $\{3,3,0\}$ & 0 & 1 &4 \\ \hline
$q^4$ & 0 & 1 & 0 & $-$ & $2\pi/M$ & $\{0,8,0\}$ & 0 & 1 & \\
      & 1 & 1 & 0 & $3\pi/M$ & $3\pi/M$ & $\{3,5,0\},\{5,3,0\}$ & 0 & 2 &6 \\
\hline
$q^5$ & 0 & 1 & 0 & $-$ & $2\pi/M$ & $\{0,10,0\}$ & 0 & 1 & \\
      & 1 & 1 & 0 & $3\pi/M$ & $3\pi/M$ & $\{3,7,0\},\{5,5,0\},\{7,3,0\}$ & 0 &
3 &8 \\ \hline
$q^6$ & 0 & 1 & 0 & $-$ & $2\pi/M$ & $\{0,12,0\}$ & 0 & 1 & \\
      & 1 & 1 & 0 & $3\pi/M$ & $3\pi/M$ & $\{3,9,0\},\{5,7,0\}$ &   &   & \\
      &   &   &   &          &          & $\{7,5,0\},\{9,3,0\}$ & 0 & 4 & \\
      & 0 & 2 & 1 & $-$ & $3\pi/M$ & $\{0,3+5,3\}$ & $\pi/M$ & 1 & 12\\ \hline
$q^7$ & 0 & 1 & 0 & $-$ & $2\pi/M$ & $\{0,14,0\}$ & 0 & 1 & \\
      & 1 & 1 & 0 & $3\pi/M$ & $3\pi/M$ & $\{3,11,0\},\{5,9,0\}$ &   &   & \\
      &   &   &   &          &          & $\{7,7,0\},\{9,5,0\},\{11,3,0\}$ & 0
&5 & \\
      & 0 & 2 & 1 & $-$ & $3\pi/M$ & $\{0,3+7,3\},\{0,3+5,5\}$ & $\pi/M$ & 2 &
\\
      & 2 & 1 & 0 & $4\pi/M$ & $4\pi/M$ & $\{4+6,4,0\}$ & 0 & 1 &18 \\ \hline
$q^8$ & 0 & 1 & 0 & $-$ & $2\pi/M$ & $\{0,16,0\}$ & 0 & 1 & \\
      & 1 & 1 & 0 & $3\pi/M$ & $3\pi/M$ & $\{3,13,0\},\{5,11,0\},\{7,9,0\}$ &
&  & \\
      &   &   &   &          &          & $\{9,7,0\},\{11,5,0\},\{13,3,0\}$ & 0
& 6 & \\
      & 0 & 2 & 1 & $-$ & $3\pi/M$ & $\{0,3+9,3\},\{0,5+7,3\}$ &  &  & \\
      &   &   &   &     &          & $\{0,3+7,5\},\{0,3+5,7\}$ &$\pi/M$ & 4 &
\\
      & 2 & 1 & 0 & $4\pi/M$ & $4\pi/M$ & $\{4+8,4,0\},\{4+6,6,0\}$ & 0 & 2 &
\\
      & 0 & 3 & 0 & $-$ & $3\pi/M$ & \{0,3+5+7,0\}& $\pi/M $ & 1 & 28\\ \hline
\end{tabular}
\end{center}
\end{table}

\setcounter{table}{6}
\begin{table}
\begin{center}
\caption{The terms through order $q^8$ in the sector
$m_{-+}=m_{++}$ with the macroscopic momentum $\Delta P=\pi$ constructed
from the rules of sec, 4.3. The minimum momenta
$P_{min}^{ns}={\pi\over M}(m_{ns}+m_{2s}+m_{-2s}+3)$ and $P_{min
}^{\pm2s}={\pi\over M}(m_{ns}+{m_{2s}+m_{-2s}+1\over 2}+3)$ are
obtained from (4.14). Here $m_{2s}+m_{-2s}$ is odd and only
$m_{2s}>m_{-2s}$ are explicitly shown.}

\begin{tabular}{|l|l|l|l|l|l|l|l|l|l|} \hline
order & $m_{ns}^\ell$ & $m_{2s}^\ell$ & $m_{-2s}^\ell$ & $P_{min}^{ns}$&
$P_{min}^{\pm2s}$&
$\{P^{ns},P^{2s},P^{-2s}\}$ (units of $\frac{\pi}{M}$)&shift &states &
total \\ \hline
$q^2$ & 0 & 1 & 0 & $-$ & $4\pi/M$ & $\{0,4,0\}$ & 0 & 1 & 2 \\ \hline
$q^3$ & 0 & 1 & 0 & $-$ & $4\pi/M$ & $\{0,6,0\}$ & 0 & 1 & 2 \\ \hline
$q^4$ & 0 & 1 & 0 & $-$ & $4\pi/M$ & $\{0,8,0\}$ & 0 & 1 & 2 \\ \hline
$q^5$ & 0 & 1 & 0 & $-$ & $4\pi/M$ & $\{0,10,0\}$ & 0 & 1 &  \\
      & 1 & 1 & 0 & $5\pi/M$ & $5\pi/M$ & $\{5,5,0\}$ & 0 & 1 &4 \\ \hline
$q^6$ & 0 & 1 & 0 & $-$ & $4\pi/M$ & $\{0,12,0\}$ & 0 & 1 &  \\
      & 1 & 1 & 0 & $5\pi/M$ & $5\pi/M$ & $\{5,7,0\},\{7,5,0\}$ & 0 & 2 &6 \\
\hline
$q^7$ & 0 & 1 & 0 & $-$ & $4\pi/M$ & $\{0,14,0\}$ & 0 & 1 & \\
      & 1 & 1 & 0 & $5\pi/M$ & $5\pi/M$ & $\{5,9,0\},\{7,7,0\},\{9,5,0\}$ & 0 &
3 &8 \\ \hline
$q^8$ & 0 & 1 & 0 & $-$ & $4\pi/M$ & $\{0,16,0\}$ & 0 & 1 & \\
      & 1 & 1 & 0 & $5\pi/M$ & $5\pi/M$ & $\{5,11,0\},\{7,9,0\}$ &  &  & \\
      &   &   &   &          &          & $\{9,7,0\},\{11,5,0\}$ &0 & 4 & \\
      & 0 & 2 & 1 & $-$ & $5\pi/M$ & $\{0,5+7,5\}$ & $-\pi/M$ & 1 & 12 \\
\hline
\end{tabular}
\end{center}
\end{table}

\setcounter{table}{7}
\begin{table}
\begin{center}
\caption{The terms through order $q^8$ in the sum of
$m_{-+}-m_{++}=1$ and the $\Delta P=0$ term of $m_{-+}=m_{++}$. These
are compared with the terms in $q^{-1/3}q^{1/24}b_0^2$ of (2.30d).}
\begin{tabular}{|c|c|c|r|r|}\hline
order & $m_{-+}-m_{++} = 1$ & $m_{-+}-m_{++} = 0$ & total & $q^{-1/3}q^{1/24}
b_0^2$ \\ \hline
$q^0$ & 1 & 0 & 1 & 1 \\
$q^1$ & 0 & 2 & 2 & 2 \\
$q^2$ & 1 & 2 & 3 & 3 \\
$q^3$ & 1 & 4 &  5 & 5 \\
$q^4$ & 2 & 6 & 8 & 8 \\
$q^5$ & 5 & 8 & 13 & 13 \\
$q^6$ & 7 & 12 & 19 & 19 \\
$q^7$ & 10 & 18 & 28 & 28 \\
$q^8$ & 13 & 28 & 41 & 41 \\ \hline
\end{tabular}
\end{center}
\end{table}
\begin{table}
\begin{center}
\caption{The terms through order $q^8$ of the sum of
$m_{-+}-m_{++}=-1$ and the $\Delta P=\pi$ term of $m_{-+}=m_{++}$.
These are compared with the terms of $q^{-1/12}q^{1/24}b_2^2$ of (2.30e)}
\begin{tabular}{|c|c|c|r|r|}\hline
order & $m_{-+}-m_{++} = -1$ & $m_{-+}-m_{++} = 0$ & total & $q^{-1/12}q^{1/24}
b_2^2$ \\ \hline
$q^0$ & 1 & 0 & 1 & 1 \\
$q^1$ & 1 & 0 & 1 & 1 \\
$q^2$ & 1 & 2 & 3 & 3 \\
$q^3$ & 2 & 2 & 4 & 4 \\
$q^4$ &6 & 2 & 8 & 8 \\
$q^5$ & 7 & 4 & 11 & 11 \\
$q^6$ & 12 & 6 & 18 & 18 \\
$q^7$ & 17 & 8 & 25 & 25 \\
$q^8$ & 26 & 12 & 38 & 38 \\ \hline
\end{tabular}
\end{center}
\end{table}

\end{document}